\documentclass[a4paper,11pt]{article}
\pdfoutput=1

\usepackage{amssymb,booktabs,comment,integrable,jheparxiv,xparse,stmaryrd}

\title{Non-Linear Schr\"{o}dinger and Korteweg--de Vries from 4d Chern--Simons Theory}

\author[a,b]{Roland Bittleston}

\affiliation[a]{Perimeter Institute for Theoretical Physics,\\ 31 Caroline Street, Waterloo, ON, Canada\vspace{0.1cm}}
\affiliation[b]{Center for the Fundamental Laws of Nature,\\ Harvard University, Cambridge, MA, USA\vspace{0.1cm}}
\emailAdd{rbittleston@fas.harvard.edu}

\begin{document}
	
\abstract{We realise the non-linear Schr\"{o}dinger and Korteweg–de Vries equations as reductions of four-dimensional Chern–Simons theory with gauge group $\mathrm{SL}_2(\mathbb{C})$, spectral curve $\mathbb{CP}^1$ and meromorphic one-form $\dif z$.  The construction uses singular boundary conditions at the double pole at infinity, together with compensating source terms that give a well-defined variational problem.  We derive the Lax connections and two-dimensional actions of both models and identify compatible real structures.  We then introduce a broader family of singular boundary conditions determined by a rank-one nilpotent endomorphism of $T^*\Sigma$.  Under suitable reality and non-degeneracy assumptions, we show that these variational problems are locally equivalent to the NLS or KdV constructions.}
	
\maketitle

\flushbottom
	

\section{Introduction} \label{sec:intro}

Over the past decade mixed holomorphic-topological 4d Chern--Simons theory, first appearing in Nekrasov's thesis \cite{Nekrasov:1996} and independently rediscovered by Costello in the context of supersymmetric gauge theory \cite{Costello:2013zra}, has emerged as a powerful organising principle for 2d integrability.  Since its introduction a vast array of models have been realised in the 4d Chern--Simons framework, ranging from quantum integrable spin chains \cite{Costello:2017dso,Costello:2018gyb} to large classes of 2d integrable field theories.  A non-exhaustive list includes the Gross--Neveu and Thirring models, the principal chiral model with and without Wess--Zumino term, and its Yang-Baxter and $\lambda$-deformations \cite{Costello:2019tri,Delduc:2019whp}.  Many more examples can be found in \cite{Lacroix:2020flf,Costello:2020lpi,Caudrelier:2020xtn,Fukushima:2021ako,Liniado:2023uoo,Lacroix:2023qlz,Cole:2024skp,Bittleston:2026tdr}.

Despite this success, a number of the most historically significant and well studied integrable partial differential equations have thus far not been directly realised in 4d Chern--Simons theory:  In particular the Korteweg--de Vries (KdV) and non-linear Schr\"{o}dinger (NLS) equations.\footnote{I was made aware of indirect realisations of quantum KdV adapting the affine Gaudin model constructions in \cite{Feigin:2007mr,Feigin:2017int} by Beno\^{\i}t Vicedo.  A comparison with our approach appears at the end of section \ref{subsec:alternatives}.}\medskip

The KdV equation for a real field $u(t,x)$ in one space and one time dimension is
\be \label{eq:KdV-equation-intro} \p_tu - \frac{1}{4}\p_x^3u - 3u\p_xu = 0\,. \ee
It describes the evolution of waves in shallow water:  $u$ represents the wave height.  The non-linear steepening term balances the dispersion, leading to soliton solutions which retain their shape over time.

The history of the KdV equation begins with John Scott Russell's celebrated discovery of a solitary wave travelling along the Union Canal in 1834.  The theory of such waves was developed in the 19\textsuperscript{th} century, notably by Boussinesq and Rayleigh.  In 1895 Korteweg and de Vries introduced the equation \eqref{eq:KdV-equation-intro} and studied its soliton and cnoidal wave solutions \cite{Korteweg:1895lrm}.  Integrability of the KdV equation was established much later via the inverse scattering transform by Gardner et al. \cite{Gardner:1967wc}, and the general principles behind the construction were subsequently elucidated by Lax, Faddeev and Zakharov, and Magri \cite{Lax:1968gpe,Faddeev:1971xd,Magri:1977gn}.

Since then the KdV equation has served as the paradigmatic integrable partial differential equation and it is ubiquitous throughout the field of mathematical physics.  Its integrability has been understood through many complementary perspectives, including Hirota's bilinear method \cite{Hirota:1971zz}, Sato theory \cite{Sato:1983sol,Jimbo:1983if,Segal:1985aga},  Drinfeld--Sokolov reduction \cite{Drinfeld:1984qv}, the topological string \cite{Douglas:1989ve,Gross:1989vs,Brezin:1990rb,Witten:1990hr,Kac:1991nv,Givental:2001gro} and at the quantum level the ODE/IM correspondence \cite{Bazhanov:1994ft,Bazhanov:1996dr,Bazhanov:1998wj,Dorey:1998pt}.\medskip

The NLS equation for a complex field $\psi(t,x)$ in one space and one time dimension is
\be \label{eq:NLS-equation-intro} \im\p_t\psi = - \frac{1}{2}\p_x^2\psi + \kappa|\psi|^2\psi \ee
for $\kappa =\pm1$.  Originally introduced in the 1960s to describe weakly interacting Bose condensates \cite{Gross:1961kqh,Pitaevskii:1961vor}, self-trapping of optical beams \cite{Chiao:1964sel} and modulation of dispersive waves \cite{Benney:1967pro,Zakharov:1968sta}, integrability of the NLS equation by the method of inverse scattering was first established by Zakharov and Shabat \cite{Zaharov:1970tma,zakharov1973interaction}.  This subsequently led to systematic development of inverse scattering as a method for solving integrable partial differential equations \cite{Ablowitz:1973zz,Ablowitz:1974ry}.  Indeed, NLS demonstrated that inverse scattering was a general construction, rather than a peculiar feature of KdV.\medskip

In this note we show how to embed the NLS and KdV equations in 4d Chern--Simons theory.  This is achieved by imposing singular boundary conditions on the dynamical partial connection, which must be compensated by suitable source terms.  We start in section \ref{sec:4dCS} by reviewing the 4d Chern--Simons framework, with an emphasis on its phase space structure.  In sections \ref{sec:NLS} and \ref{sec:KdV} we explain how to recover NLS and KdV respectively.  In both cases we begin by reviewing the Lax and variational principle.  We then introduce our singular boundary conditions and the source terms required to compensate them.  Considerable space is devoted to gauge fixing and imposing the fibre equations of motion, followed by the compactification to 2d.  We finish by listing compatible real structures.  In section \ref{sec:gen} we show that the NLS and KdV boundary conditions belong to a family depending on a rank one nilpotent map $N:T^*\Sigma\to T^*\Sigma$.  Under suitable non-degeneracy assumptions, we show that any boundary condition in this family is, through a combination of co-ordinate and gauge transformations, locally equivalent to the NLS or KdV case.


\subsection{Alternative Organising Principles} \label{subsec:alternatives}

Let us briefly discuss two alternative unifying frameworks for integrable systems:  Ward's conjecture concerning symmetry reductions of the self-dual Yang--Mills equations and affine Gaudin models.  Both are closely related to the 4d Chern--Simons framework.

Ward's conjecture states that `many (and perhaps all?)\ of the partial differential equations that are regarded as being integrable or solvable may be obtained from the self-dual gauge field equations (or its generalisations) by reduction' \cite{Ward:1985gz}, where a reduction refers to the condition of equivariance of the gauge connection under a subgroup of the conformal group lifted to the gauge bundle.  The conjecture is supported by a remarkable range of examples:  Even at the time of its proposal known reductions included the Bogomoln'yi equations in three dimensions, Toda theory \cite{Leznov:1980tz} and the integrable chiral model in two dimensions and the Nahm equations in one dimension.  Since then the list has expanded to include Hitchin's equations \cite{Hitchin:1986vp} and the KdV and NLS equations \cite{Mason:1989kk,Mason:1992vd}.  Even the Kadomtsev--Petviashvili equation, which does not appear to be recoverable in the strictest sense, can be obtained by introducing space-time non-commutativity \cite{Strachan:1994kp,Strachan:1994tx} or using infinite dimensional gauge algebras \cite{Chakravarty:1995red}.  Systematic accounts of those integrable equations which arise as symmetry reductions of the self-dual Yang--Mills equations can be found in \cite{Mason:1996rf,Ablowitz:2003bv}.

The self-dual Yang--Mills equations can be formally solved using twistor theory.  Indeed, self-dual gauge connections are in correspondence with holomorphic vector bundles on twistor space, which are characterised by algebraic data \cite{Ward:1977ta}.  Given a reduction of the self-dual Yang--Mills equations, the corresponding subgroup of the conformal group $H$ lifts to the left-handed projective spin bundle $\PS = \bbR^4\times S^2_\mrm{L}$, and identifying $\PS$ with Euclidean twistor space $\PT$ it also inherits an action of the complexification $H_\bbC$.  The quotient $\PT/H_\bbC$ is a `micro-twistor space', and the quotient $\PS/H$ gives a `micro-twistor correspondence space'.  For a locally free two-dimensional reduction $\PT/H_\bbC$ has complex dimension one and $\PS/H$ has real dimension four.  In this way the twistor correspondence induces `micro-twistor correspondences' on its reductions \cite{Mason:1996rf}.  In \cite{Bittleston:2020hfv} it was argued that the domain of 4d Chern--Simons theory can naturally be viewed as a micro-twistor correspondence space, and furthermore that the 4d Chern--Simons action could be obtained as a symmetry reduction of holomorphic Chern--Simons theory on twistor space.  This perspective has since been used to realise a number of 2d integrable models in 4d Chern--Simons theory \cite{Cole:2023umd,Cole:2024sje,Jarov:2025bmq,Ashwinkumar:2026zrr}.

Given that both the KdV and NLS equations could be obtained by symmetry reduction from the self-dual Yang--Mills equations, it's natural to expect that they can also be embedded in 4d Chern--Simons.  Furthermore, the Kadomtsev--Petviashvili equation can be formulated as a non-commutative 5d Chern--Simons theory on mini-twistor correspondence space \cite{Bittleston:2025gxr}, which should also symmetry reduce to a 4d Chern--Simons description of KdV.  Although the boundary conditions introduced in this note were not engineered via symmetry reduction, this perspective strongly motivated their form.\medskip

On the other hand, Gaudin models describe a collection of sites each labelled by a representation of a fixed simple Lie algebra and a spectral parameter valued in a fixed Riemann surface.  The Hamiltonian is a sum of interactions between pairs of sites, with interaction strength depending on the corresponding pair of spectral parameters \cite{Gaudin:1976dia,Gaudin:2014bet}.  In an affine Gaudin model the simple Lie algebra is replaced by an affine Lie algebra \cite{Feigin:2007mr}, yielding models with an infinite number of degrees of freedom.  Many integrable field theories can be obtained as instances of affine Gaudin models and variants thereof \cite{Vicedo:2017cge,Lacroix:2023gig}.  At least for standard boundary conditions, Hamiltonian analysis of 4d Chern--Simons theory yields affine Gaudin models \cite{Vicedo:2019dej}.

Whilst we're not aware of an explicit rendering of KdV in 4d Chern--Simons theory, I learnt from Beno\^{\i}t Vicedo that quantum KdV can be realised through an affine Gaudin model in at least two different ways.  Translated into 4d Chern--Simons theory, the first \cite{Feigin:2007mr} (Section 5) uses
\be \omega = k\frac{\dif z}{z} \ee
and the second \cite{Feigin:2017int} (Section 7.3)
\be \omega = \bigg(\frac{k}{z} + \frac{1}{z-1}\bigg)\dif z \ee
for $k\in\bbZ_{>0}$.  Both constructions engineer the Virasoro vertex algebra (at different central charges); time evolution by the Hamiltonian $I_3 \sim \int\normord{T^2}$ recovers quantum KdV.

It would be very interesting to relate the construction of KdV in this note to affine Gaudin models through a Hamiltonian analysis, and specifically to the above two realisations.  Superficially the two approaches seem inequivalent:  The meromorphic one-form $\omega$, and hence the corresponding boundary conditions, differ.  Furthermore, the 4d Chern--Simons formulation presented in section \ref{sec:KdV} recovers the first Poisson structure and Hamiltonian $I_5 = \int u^3 - \tfrac{1}{4}(\p_xu)^2$, rather than the second Poisson structure and Hamiltonian $I_3$ more naturally arising in the classical limit of Virasoro.


\section{Mixed Holomorphic-Topological 4d Chern--Simons Theory} \label{sec:4dCS}

Here we recall the action, equations of motion and phase space symplectic structure of 4d Chern--Simons, following \cite{Costello:2019tri}.  We also review how 2d integrable field theories are realised in this framework.

The theory is defined over a 4d manifold with the product structure $M = \Sigma\times C$ for $\Sigma$ a smooth 2d manifold and $C$ a Riemann surface equipped with a meromorphic one-form $\omega$.\footnote{More generally it can be defined on 4d manifolds which are transversally holomorphically fibred over Riemann surfaces equipped with a meromorphic one-form.}  We use the real co-ordinates $t,x$ on $\Sigma$ and the holomorphic co-ordinate $z$ on $C$.  The theory also depends on a choice of Lie group $G$ with Lie algebra $\fg$ and invariant bilinear form $\tr$.  We will always take the gauge bundle to be trivial $P = G\times M$.

The dynamical field of 4d Chern-Simons theory is a partial connection
\be a = a_t\dif t + a_x\dif x + a_{\bar z}\dif\bar z \in \widetilde\Omega^{0,1}(M;\fg)\,, \ee
where $\widetilde\Omega^{0,q}(M;\fg)$ is the vector space of smoothly varying differential $q$-forms on $M$ modulo $\dif z$.  We can (and will) always choose representatives of this quotient built solely from wedge products of $\dif t,\dif x$ and $\dif\bar z$.  The partial connection $a$ has an associated gauge symmetry parametrised infinitesimally by $\eps\in\widetilde\Omega^{0,0}(M;\fg) = C^\infty(M;\fg)$ and acting by
\be \delta_\eps a = \dift\eps + [a,\eps]\,, \ee
where
\be \dift = \dif_\Sigma + \dbar_C = \dif t\p_t + \dif x\p_x + \dif\bar z\p_{\bar z}\,. \ee
The 4d Chern--Simons action is
\be \label{eq:4dCS-bulk} S_\text{bulk}[a] = \frac{\im}{2\pi}\int_M\omega\wedge\CS(a) \ee
where we abuse notation by confusing $\omega$ with its pullback to $M$, and
\be \CS(a) = \tr\bigg(a\wedge\dift a + \frac{1}{3}a\wedge[a,a]\bigg) \ee
is the mixed holomorphic-topological Chern--Simons three-form.  We refer to \eqref{eq:4dCS-bulk} as the bulk action for 4d Chern--Simons theory, as in the sequel it needs to be improved by source terms.

In general $\omega$ will have both poles and zeros in $C$, potentially of degree greater than one.  In a neighbourhood of the corresponding loci in $M$, the gauge field $a$ requires careful treatment:
\begin{itemize}
	\renewcommand{\labelitemi}{$\blacktriangleright$}
	\item  Around poles of $\omega$ constraints need to be imposed on $a$ to ensure that the action is finite and the variational problem is well defined.  Gauge transformations are similarly restricted to ensure that they preserve these constraints.
	\item  Around zeros of $\omega$ smoothness of $a$ must be relaxed to ensure that the equations of motion have non-trivial solutions.  In particular, components of $a$ are permitted to have poles.
\end{itemize}
It's also possible to engineer more elaborate conditions which combine poles and zeros in different components of $a$, but always in such a way as to ensure that the action remains finite and the equations of motion have non-trivial solutions.

To understand the 4d Chern--Simons integrand in neighbourhoods of the zeros and poles of $\omega$ we find it useful to regulate the domain $M$.  In particular we excise small discs of radius $\eps$ around the poles and zeros of $\omega$ in $C$, yielding a Riemann surface with boundary $C_\eps$ on which $\omega$ is holomorphic and non-vanishing.  The regulated domain is the 4d manifold with boundary $M_\eps = \Sigma\times C_\eps$.   We can then evaluate the action on $M_\eps$, and any boundary contributions on $\p M_\eps$, before taking the limit $\eps\to0$.  Indeed, from now on we will refer to the constraints obeyed or relaxations enjoyed by $a$ at poles and zeros as `boundary conditions', despite the fact that in the $\eps\to0$ limit these constraints are imposed at co-dimension 2 loci.

Varying the regulated 4d Chern--Simons action gives
\bea \label{eq:4dCS-variation}
&\delta S_\text{bulk}[a] = 
- \frac{\im}{\pi}\lim_{\eps\to0}\int_{M_\eps}\omega\wedge\tr\big(\delta a\wedge f(a)\big) - \frac{\im}{2\pi}\lim_{\eps\to0}\int_{\p M_\eps}\omega\wedge\tr(a\wedge\delta a)\,.
\eea
where $f(a) = \dift a + \tfrac{1}{2}[a,a]$ and $\delta$ is the de Rham differential on the configuration space of gauge fields, and is therefore graded odd.  The bulk equations of motion are
\be \omega\wedge f(a) = 0\,. \ee
In the gauge $a_{\bar z} = 0$ the equations of motion read
\be \p_{\bar z}a_t = \p_{\bar z}a_x = 0\,,\qquad \p_ta_x - \p_xa_t + [a_t,a_x] = 0\,, \ee
so that $a$ is a flat connection depending meromorphically on a spectral parameter $z$.  In this way we can interpret the partial connection $a$ as a dynamical analogue of the Lax connection of a 2d integrable model.  In practice the gauge $a_{\bar z} = 0$ is obstructed, but this is nevertheless a good heuristic.

We can also read off the phase space pre-symplectic potential from \eqref{eq:4dCS-variation}
\be \label{eq:symplectic-potential} \Theta = \frac{\im}{2\pi}\lim_{\eps\to0}\int_{-\p M_\eps}\omega\wedge\tr(a\wedge\delta a)\,. \ee
Here we've opted to reverse the orientation on $\p M_\eps$ so that boundary circles of $C_\eps$ are traversed in the standard anti-clockwise orientation.  The phase space pre-symplectic structure is therefore
\be \label{eq:symplectic-form} \Omega = \delta\Theta = \frac{\im}{2\pi}\lim_{\eps\to0}\int_{-\p M_\eps}\omega\wedge\tr\big(\delta a\wedge\delta a\big)\,, \ee
where we've taken care to note that $\delta$ anti-commutes with integration over odd-dimensional manifolds.

Boundary conditions are Lagrangian submanifolds of the boundary phase space; the space of solutions to the equations of motion in a neighbourhood of the boundary locus.  In a gauge theory it's often convenient to replace this phase space by its derived counterpart, which retains the gauge symmetry and constraint data.  The natural mathematical language for describing such derived quotients is the Batalin–-Fradkin–-Vilkovisky (BFV) formalism.  In this setting boundary conditions are Lagrangians in the derived boundary phase space.  We will not utilise the BFV formalism in this note, but our boundary conditions will be chosen so that they determine a Lagrangian in this derived sense.

Our 4d Chern--Simons boundary conditions will restrict the dynamical field $a$ and gauge transformations $\eps$.  They are required to obey the following compatibility conditions:
\begin{itemize}
	\renewcommand{\labelitemi}{$\blacktriangleright$}
	\item  The allowed gauge transformation should close onto allowed gauge transformations.
	\item  The allowed gauge transformations should preserve the boundary conditions on $a$.
	\item  The boundary conditions on $a$ should be Lagrangian for the pre-symplectic structure \eqref{eq:symplectic-form}.  That is, they should be isotropic and co-isotropic.  The latter condition ensures that the boundary conditions are not `too strong', in particular that they are first class. 
\end{itemize}
There is one further requirement:  The pre-symplectic potential \eqref{eq:symplectic-potential} may not be zero when boundary conditions are imposed, but it will be $\delta$ closed.  If it is $\delta$ exact then the action can be improved by source terms, correcting the pre-symplectic potential so that it too vanishes.  Our boundary conditions will be chosen so that a suitable improvement exists.

In the bulk of this paper we take $C = \CP^1$ with stereographic co-ordinate $z$.  For the meromorphic one-form we choose
\be \omega = \dif z\,, \ee
which has a double pole at $z=\infty$.  This pole is visible in the patch with co-ordinate $w = - 1/z$ in terms of which
\be \omega = \frac{\dif w}{w^2}\,. \ee
We will also take $\fg = \fsl_2(\bbC)$ with $\tr$ the fundamental trace.


\subsection{2d Integrable Field Theories} \label{subsec:integrable-models}

There are broadly speaking two different methods of obtaining 2d integrable field theories from 4d Chern-Simons, through the introduction of order or disorder defects:

\paragraph{Order defects.}  4d Chern--Simons theory is coupled to 2d surface defects wrapping $\Sigma$ and supported at points of $C$ which carry independent degrees of freedom through some $G$ action.  Integrating out the field $a$ couples such defects together, generating a 2d integrable model.  Examples of models which can be obtained this way include the Gross--Neveu and massless Thirring models, as well as $\sigma$-models with K\"{a}hler targets.

\paragraph{Disorder defects.}  The diffeomorphism invariance of 4d Chern--Simons theory is explicitly broken by boundary conditions on $a$ at poles and zeros of $\omega$.  This necessitates restrictions on gauge transformations, so that the gauge equivalence classes of $\p_{\bar z} + a_{\bar z}$ on the fibre $C_{(t,x)}$ at fixed $(t,x)\in\Sigma$ form a non-trivial moduli space $\cM(C;G)$.  Compactification to 2d is achieved by integrating out fluctuations of $a_\Sigma$ (or classically imposing the corresponding equation of motion), resulting in a 2d $\sigma$ model valued in $\cM(C;G)$.  Examples of models which can be obtained this way include the principal chiral model, both with and without Wess--Zumino term, together with its integrable deformations.\medskip

In this work we will be concerned with the latter approach.  In both cases integrability of the resulting 2d field theory is manifest in the 4d Chern--Simons description.  In brief:  The partial connection $a$ is fixed, up to a (potentially boundary condition violating) gauge transformation, to the Lax.  Compactifying the $x$ direction to $S^1$, Wilson loops around the compact cycle yield the transfer matrix
\be T_\mrm{R}(t;z) = \tr_\mrm{R}\mrm{P}\exp\bigg( - \oint_{S^1}a_x\bigg)\,. \ee
Diffeomorphism invariance of 4d Chern--Simons in the $\Sigma$ direction, which is retained away from poles and zeros of $\omega$, ensures time independence and Poisson commutativity of transfer matrices.\footnote{This appealing picture can be misleading:  In non-ultralocal theories (which arise when $\omega$ has zeros) the Poisson bracket of transfer matrices can be ill-defined.  Nevertheless, this argument gives an intuition for how integrability emerges from 4d Chern--Simons theory.}  Expanding the $T_\mrm{R}(z)$ either as a Laurent series or Taylor series in the spectral parameter around poles and zeros of $\omega$ gives infinite families of conserved charges.


\subsection{Regulating Boundary Terms} \label{subsec:eps-reg}

Throughout this work we will encounter contour integrals of the form
\be \label{eq:example-singularity} \Res_{w=0}\frac{\dif w}{w^k}F(w,\bar w) \coloneqq \frac{1}{2\pi\im}\lim_{\eps\to0}\oint_{|w|=\eps}\frac{\dif w}{w^k}F(w,\bar w) \ee
for $F$ a smooth function on a disc $D^2_R$ of radius $R>\eps$ centred at $w=0$ and $k$ a positive integer.  Crucially it need not be that $F$ vanishes at $w=0$, so the integrand is singular in the $\eps\to0$ limit.  Nevertheless, the integral remains finite and evaluates to
\be \Res_{w=0}\frac{\dif w}{w^k}F(w,\bar w) = \frac{1}{(k-1)!}\p_w^{k-1}F(0)\,. \ee
For example, a monomial $F(w,\bar w) = w^m\bar w^n$ can only contribute if $m - n = k-1$, in which case the integral evaluates to $\lim_{\eps\to0}\eps^{2n} = \delta_{n=0}$.


\section{Non-Linear Schr\"{o}dinger} \label{sec:NLS}

We begin by concentrating on the non-linear Schr\"{o}dinger (NLS) equation
\be \label{eq:NLS-equation} \im\p_t\psi = - \frac{1}{2}\p_x^2\psi + \kappa|\psi|^2\psi \ee
for a complex valued wavefunction $\psi\in C^\infty(\Sigma;\bbC)$.  Up to rescaling of the dependent and independent variables we can fix $\kappa = \pm 1$.  The case $\kappa = -1$ is known as attractive or focussing, whereas $\kappa = 1$ is repulsive or anti-focussing.  The NLS equation is integrable with Lax connection
\bea \label{eq:NLS-Lax}
&\cL_t - z\cL_x = \frac{\im}{2}\begin{pmatrix} \kappa|\psi|^2 & \p_x\psi \\ - \kappa\p_x\bar\psi\ & - \kappa|\psi|^2 \end{pmatrix}\,,\qquad \cL_x = \begin{pmatrix} \im z & \psi \\ \kappa\bar\psi & - \im z \end{pmatrix}\,.
\eea
A variational formulation is supplied by the action
\be \label{eq:NLS-action} S_\text{NLS}[\psi,\bar\psi] = \int_\Sigma\dif t\wedge\dif x\,\bigg(\im\bar\psi\p_t\psi - \frac{1}{2}|\p_x\psi|^2 - \frac{1}{2}\kappa|\psi|^4\bigg)\,. \ee

For the majority of this section we'll work with the complexified non-linear Schr\"{o}dinger equation, which has two independent complex valued fields $\psi,\tilde\psi\in C^\infty(\Sigma;\bbC)$ obeying
\be \label{eq:NLS-equation-complex} \im\p_t\psi = - \frac{1}{2}\p_x^2\psi + \tilde\psi\psi^2\,,\qquad -\im\p_t\tilde\psi = - \frac{1}{2}\p_x^2\tilde\psi + \tilde\psi^2\psi\,. \ee
These are obtained by varying the complex action
\be \label{eq:NLS-action-complex} S_{\bbC\text{NLS}}[\psi,\tilde\psi] = \int_\Sigma\dif t\wedge\dif x\,\bigg(- \im\tilde\psi\p_t\psi + \frac{1}{2}\p_x\tilde\psi\p_x\psi + \frac{1}{2}\tilde\psi^2\psi^2\bigg)\,. \ee

Here we'll see that the NLS equation can be obtained from 4d Chern--Simons theory.  In section \ref{subsec:NLS-bcs} we introduce boundary conditions for the fields and gauge transformations of 4d Chern--Simons which lead to NLS.  These boundary conditions are singular, and require the introduction of source terms supported at the pole in $\omega$, which we determine in section \ref{subsec:NLS-source}.  We show in section \ref{subsec:NLS-Lax} that up to gauge the partial connection is determined by a pair of complex-valued fields on $\Sigma$ which we identify with $\psi,\tilde\psi$.  Further we argue that imposing the equations of motion with one component in the $\CP^1$ direction fixes $a$ to be gauge equivalent to the Lax of the non-linear Schr\"{o}dinger equation \eqref{eq:NLS-Lax}.  In section \ref{subsec:NLS-compactify} we evaluate the 4d Chern--Simons action on this configuration to recover \eqref{eq:NLS-action-complex}.  The attractive and repulsive non-linear Schr\"{o}dinger equations are recovered by fixing the real structures
\be \tilde\psi = - \bar\psi\qquad\text{and}\qquad\tilde\psi = \bar\psi \ee
respectively.  We explain how to recover these from an involutive automorphism of 4d Chern--Simons theory in section \ref{subsec:NLS-reality}.


\subsection{Boundary Conditions} \label{subsec:NLS-bcs}

We will take $C = \CP^1$ and the meromorphic one-form to be
\be \omega = \dif z = \frac{\dif w}{w^2}\,, \ee
and use gauge group $\gSL_2(\bbC)$.  We regulate as in section \ref{sec:4dCS} by excising a disc of small radius $\eps>0$ around $w = 0$, before taking $\eps\to0$.  Neglecting boundaries in $\Sigma$ the pre-symplectic structure \eqref{eq:symplectic-form} specialises to
\be \label{eq:NLS-symplectic} \Omega = - \int_\Sigma\Res_{w=0}\frac{\dif w}{w^2}\wedge\tr\big(\delta a\wedge\delta a\big) = 2\int_\Sigma\dif t\wedge\dif x\wedge\Res_{w=0}\frac{\dif w}{w^2}\wedge\tr\big(\delta a_t\wedge\delta a_x\big)\,, 
\ee
where $\Res$ is defined in equation \eqref{eq:example-singularity}.

We will use the basis $\{h,e,f\}$ of $\fsl_2(\bbC)$ with commutators
\be [h,e] = 2e\,,\qquad [h,f] = -2f\,,\qquad [e,f] = h \ee
and invariant symmetric bilinear $\tr(h^2) = 2$ and $\tr(ef) = 1$ with all other matrix elements zero.  Our boundary conditions will break global $\gSL_2(\bbC)$ invariance, and it will be useful to introduce the components
\be a = a^h h + a^e e + a^f f \ee
in terms of which the pre-symplectic form is
\be \Omega = 2\int_\Sigma\dif t\wedge\dif x\wedge\Res_{w=0}\frac{\dif w}{w^2}\wedge\big(\delta a_t^e\wedge\delta a_x^f + \delta a_t^f\wedge\delta a_x^e + 2\delta a_t^h\wedge\delta a_x^h\big)\,. \ee
Our boundary conditions on $a$ will need to be isotropic for this form.

We prescribe fixed singular behaviour on the space-time components of the gauge field as $w\to0$, and impose restrictions on fluctuations around this background.  That is, we write
\be \label{eq:NL-split} a = a_0 + b \ee
where
\be \label{eq:NLS-a0} a_0 = z\im h(z\dif t + \dif x) = \frac{1}{w^2}\im h(\dif t - w\dif x) \ee
solves the 4d Chern--Simons equations of motion away from $\Sigma_0$, and $b$ is non-singular.  We further require that the space-time components of $b$ obey the boundary conditions
\bea \label{eq:NLS-bcs}
&wb^e_t + b^e_x  = \mrm{O}(w^2)\,,\qquad wb^f_t + b^f_x  = \mrm{O}(w^2)\,,\qquad b_t^h = \mrm{O}(w)\,, \\
&b_x^e = \mrm{O}(w)\,,\qquad b_x^f = \mrm{O}(w)\,,\qquad b_x^h = \mrm{O}(w)\quad\text{as}\quad w\to0\,,
\eea
where the notation $F = \mrm{O}(w^n)$ indicates that $F = w^nG$ for a smooth function $G\in C^\infty(M)$.  The last three conditions can collectively be written as $b_x = \mrm{O}(w)$ as $w\to0$.  Since $a_0$ is fixed the variations of $a$ and $b$ coincide: $\delta a = \delta b$.  The boundary conditions \eqref{eq:NLS-bcs} ensure that
\be \label{eq:NLS-quadratic-vanishing} \tr(\delta a_t\wedge\delta a_x) =  \delta b^e_t\wedge\delta b^f_x + \delta b^f_t\wedge\delta b^e_x + 2\delta b^h_t\wedge\delta b^h_x = \mrm{O}(w^2) \ee
as $w\to0$.  For the $h$ components this is immediate, for the $e$ and $f$ components
\be \delta b^e_t\wedge \delta b^f_x + \delta b^f_t\wedge \delta b^e_x = - w\delta b^e_t\wedge \delta b^f_t - w\delta b^f_t\wedge \delta b^e_t + \mrm{O}(w^2) = \mrm{O}(w^2)\,, \ee
where in the first equality we've used the identities $wb^e_t + b^e_x  = \mrm{O}(w^2)$, $wb^f_t + b^f_x  = \mrm{O}(w^2)$ and in the second we've noted that $\delta b^e_t,\delta b^f_t$ are one-forms and so anti-commute.  The second order vanishing in equation \eqref{eq:NLS-quadratic-vanishing} as $w\to0$ guarantees that the boundary conditions \eqref{eq:NLS-bcs} are isotropic for the pre-symplectic structure \eqref{eq:NLS-symplectic}.

Boundary conditions of this form necessitate the introduction of source terms in the Lagrangian localised at $w = 0$.  These will be treated carefully in section \ref{subsec:NLS-source}, but for the time being we'll trust that prescribing singularities in $a$ of the form given in equation \eqref{eq:NLS-a0} is a legal move.

Next we need to check whether the boundary conditions \eqref{eq:NLS-bcs} are co-isotropic: This requires determining for which $\xi\in\widetilde\Omega^{0,1}(M;\fsl_2(\bbC))$ the pairing
\bea
&\Omega(\xi,b) = 2\int_\Sigma\dif t\wedge\dif x\Res_{w=0}\frac{\dif w}{w^2}\big(\xi_t^e(wb_t^f + b_x^f) + \xi_t^f(wb_t^e + b_x^e) + 2\xi_t^hb_x^h \\
& - b_t^e(w\xi_t^f + \xi_x^f) - b_t^f(w\xi_t^e + \xi_x^e) - 2b_t^h\xi_x^h\big)
\eea
is zero for any $b$ obeying the boundary conditions.  For the residue to vanish, the term in brackets needs to be $\mrm{O}(w^2)$ as $w\to0$.  We infer that
\bea \label{eq:NLS-symplectic-conjguate}
&\xi_t^e = \mrm{O}(1)\,,\qquad \xi_t^f = \mrm{O}(1)\,,\qquad\xi_t^h = \mrm{O}(w)\,, \\
&w\xi_t^e + \xi_x^e = \mrm{O}(w^2)\,,\qquad w\xi_t^f + \xi_x^f = \mrm{O}(w^2)\,,\qquad\xi_x^h = \mrm{O}(w)\,,\quad\text{as}\quad w\to0\,.
\eea
Notice that $\xi_t^e = \mrm{O}(1)$ is the statement that $\xi_t^e$ is non-singular as $w\to0$; together with $w\xi_t^e + \xi_x^e = \mrm{O}(w^2)$ it implies that $\xi_x^e = \mrm{O}(w)$.  Similarly $\xi_x^f = \mrm{O}(w)$.  The restrictions \eqref{eq:NLS-symplectic-conjguate} therefore imply that $\xi$ obeys the boundary conditions \eqref{eq:NLS-bcs} establishing the Lagrangian condition and in particular co-isotropy.

Although we've fixed boundary conditions on the space-time components $a$, we must further impose compatible boundary conditions on the remaining component $a_{\bar w}$ and gauge transformations.  Recall that an infinitesimal gauge transformation with parameter $\eps\in C^\infty(M;\fsl_2(\bbC))$ acts by
\be \delta_\eps a = \delta_\eps b = \dift\eps + [a_0,\eps] + [b,\eps]\,. \ee
We must restrict $\eps$ so that the boundary conditions \eqref{eq:NLS-bcs} are preserved.

\begin{claim}
	Infinitesimal gauge transformations $\eps = \eps^ee + \eps^ff + \eps^hh$ obeying
	\be \label{eq:NLS-eps-bcs} \eps^e = \mrm{O}(w^2)\,,\qquad\eps^f = \mrm{O}(w^2)\,,\qquad \eps^h = \mrm{O}(w)\quad\text{as}\quad w\to0 \ee
	respect the boundary conditions \eqref{eq:NLS-bcs}.
\end{claim}

\begin{proof}
	Since $\eps = \mrm{O}(w)$ and $b$ is non-singular the terms $\dift\eps$ and $[b,\eps]$ respect all but $wb^e_t + b^e_x  = \mrm{O}(w^2)$ and $wb^f_t + b^f_x  = \mrm{O}(w^2)$.  The stronger constraints $\eps^e = \eps^f = \mrm{O}(w^2)$ ensure that the $\dift\eps$ term also respects these conditions.  Furthermore
	\bea
	&[wb_t + b_x,\eps]^e = 2(wb_t^h + b_x^h)\eps^e - 2(wb_t^e + b_x^e)\eps^h = \mrm{O}(w^3)\,, \\
	&[wb_t + b_x,\eps]^f = 2(wb_t^f + b_x^f)\eps^h - 2(wb_t^h + b_x^h)\eps^f = \mrm{O}(w^3)\,,
	\eea
	for example, $wb_t^e + b_x^e$ is $\mrm{O}(w^2)$ and $\eps^h$ is $\mrm{O}(w)$.
	
	This leaves the background term.  The constraints \eqref{eq:NLS-eps-bcs} imply that $\eps = wHh + w^2S$ for $H\in C^\infty(M)$ and $S\in C^\infty(M;\fsl_2(\bbC))$, so that
	\bea
	&[a_{0,t},\eps] = [\im h,S] = 2\im eS^e - 2\im fS^f\,, \\
	&[a_{0,x},\eps] = - w[\im h,S] = - 2\im w eS^e + 2\im wfS^f\,.
	\eea
	The right-hand sides are consistent with the boundary conditions \eqref{eq:NLS-bcs}.
\end{proof}

We can understand these boundary conditions more abstractly by defining a Lie subalgebra $\cI_\fh$ of $C^\infty(M;\fg)$.  $\cI_\fh$ consists of all smooth maps $\eta\in C^\infty(M;\fg)$ obeying
\be \eta^e = \mrm{O}(w^2)\,,\qquad\eta^f = \mrm{O}(w^2)\,,\qquad\eta^h = \mrm{O}(w)\quad\text{as}\quad w\to0\,. \ee
Our boundary conditions on infinitesimal gauge transformations state that $\eps\in\cI_\fh$.  Crucially $[\cI_\fh,\cI_\fh]\subset \cI_\fh$, ensuring that gauge transformations close.  The finite counterpart of an infinitesimal gauge transformation is a group valued map $k:M\to\gSL_2(\bbC)$ acting on the partial connection $a$ by
\be \label{eq:NLS-finite-gauge} a_0 + b \mapsto k\dift k^{-1} + k(a_0 + b)k^{-1}\,. \ee
The boundary conditions on infinitesimal gauge transformations \eqref{eq:NLS-eps-bcs} translate into the condition $\log k\in\cI_\fh$ in a neighbourhood of $w=0$.

Finally we need to impose boundary conditions on the $\dif\bar w$ components of $a$.  We will simply require that $a_{\bar w}\in\cI_\fh$, i.e.,
\be \label{eq:NLS-wbar-bcs} a_{\bar w}^e = \mrm{O}(w^2)\,,\qquad a_{\bar w}^f = \mrm{O}(w^2)\,,\qquad a_{\bar w}^h = \mrm{O}(w)\quad\text{as}\quad w\to0 \ee
 which is a consistent boundary condition since $\p_{\bar w}\cI_\fh\subset\cI_\fh$ and $[\cI_\fh,\cI_\fh]\subset \cI_\fh$.  These boundary conditions ensure that, in a neighbourhood of $w=0$, $a_{\bar w}$ can always be eliminated by an allowed gauge transformation.  There is no independent co-isotropy condition on $a_{\bar w}$, since this component does not enter the spectral boundary two-form.\footnote{In the formal holomorphic model near $w=0$, our boundary conditions extend to a Lagrangian differential graded Lie subalgebra $\mathfrak{l}\subset\mathfrak{b} = \Omega^\bullet(\Sigma;\fsl_2(\bbC)(\!(w)\!))$ with differential $\Dif_{0,\Sigma} = \dif_\Sigma + \ad_{a_0}$.  The residue pairing \eqref{eq:NLS-symplectic}, extended to polyforms, induces a degree-zero symplectic pairing on $\mathfrak{b}[1]$.  The zero- and one-forms in $\mathfrak{l}$ obey the boundary conditions \eqref{eq:NLS-eps-bcs} and \eqref{eq:NLS-bcs} respectively; its two-forms lie in the annihilator of its zero-forms with respect to the residue pairing.  Both $\mathfrak{b}$ and $\mathfrak{l}$ are quasi-isomorphic to their Dolbeault resolutions.  In the latter, the $\bar w$ component of the one-form field is valued in the degree zero part of $\mathfrak{l}$, giving the boundary condition \eqref{eq:NLS-wbar-bcs}.}

Away from $\Sigma_0 = \{w=0\}\subset M$ the background connection $a_0$ obeys the 4d Chern--Simons equations of motion, but on the locus $\Sigma_0$ it's sourced.  We now need to determine the corresponding source term.


\subsection{Source Terms} \label{subsec:NLS-source}

We can determine the source terms which must be adjoined to the action by finding a potential for the pre-symplectic potential \eqref{eq:symplectic-potential} when restricted to the boundary conditions of the previous section.  Recall that the pre-symplectic potential is
\be
\Theta = - \int_\Sigma\mrm{Res}_{w=0}\frac{\dif w}{w^2}\wedge\tr(a\wedge \delta a) = \int_\Sigma\dif t\wedge\dif x\,\Res_{w=0}\frac{\dif w}{w^2}\,\tr\big(a_t\delta a_x - a_x\delta a_t\big)\,.
\ee
For our pinned boundary conditions it restricts to
\bea
&\Theta|_\text{bc} = - \int_\Sigma\mrm{Res}_{w=0}\frac{\dif w}{w^2}\wedge\tr(a_0\wedge \delta a) = \int_\Sigma\dif t\wedge\dif x\,\mrm{Res}_{w=0}\frac{\dif w}{w^4}\tr\big(\im h(\delta a_x + w\delta a_t)\big)\,,
\eea
which can be compensated by adding the source term
\bea \label{eq:NLS-source}
&S_\text{source}[b;a_0] = - \int_\Sigma\mrm{Res}_{w=0}\frac{\dif w}{w^2}\wedge\tr(a_0\wedge b) \\
&= - \int_\Sigma\dif t\wedge\dif x\,\mrm{Res}_{w=0}\frac{\dif w}{w^4}\tr\big(\im h(b_x + wb_t)\big)
\eea
to the action.\footnote{It's perhaps more accurate to refer to this as a boundary term; however, we'll see shortly that the bulk 4d Chern--Simons functional localises to the boundary so we prefer to distinguish it as a source.}  Notice that shifting $b\mapsto a = a_0 + b$ in the argument does not change the source term, since $a_0\wedge a_0 = 0$.  

\begin{claim}
	The boundary term \eqref{eq:NLS-source} is invariant under the infinitesimal gauge transformations \eqref{eq:NLS-eps-bcs}.
\end{claim}

\begin{proof}
Under an infinitesimal gauge transformation
\be
\delta_\eps S_\text{source}[b;a_0] = - \int_\Sigma\mrm{Res}_{w=0}\frac{\dif w}{w^2}\wedge\tr\big(a_0\wedge\dif_\Sigma\eps + a_0\wedge[a_0,\eps] + a_0\wedge[b,\eps]\big)\,. \ee
The first term in the trace is exact on $\Sigma$ and the second vanishes because $[a_0,a_0] = 0$.  This leaves
\bea \label{eq:NLS-gauge-Sbdry}
&\delta_\eps S_\text{source}[b;a_0] = - \int_\Sigma\mrm{Res}_{w=0}\frac{\dif w}{w^2}\wedge\tr\big(a_0\wedge[b,\eps]\big) \\
&= - 2\im\int_\Sigma\dif t\wedge\dif x\Res_{w=0}\frac{\dif w}{w^4}[b_x + wb_t,\eps]^h
\eea
Writing out the commutator
\bea
&[b_x + wb_t,\eps]^h = (b_x^e + wb_t^e)\eps^f - (b_x^f + wb_t^f)\eps^e\,.
\eea
The boundary conditions \eqref{eq:NLS-bcs} and \eqref{eq:NLS-eps-bcs} imply that the right-hand side is $\mrm{O}(w^4)$, for example, $b_x^e + wb_t^e = \mrm{O}(w^2)$ and $\eps^f = \mrm{O}(w^2)$.  Hence the gauge variation of the boundary term vanishes.
\end{proof}
Let us write $S_\text{4dCS}[a;a_0] = S_\text{bulk}[a] + S_\text{source}[b;a_0]$ for the 4d Chern--Simons action including its boundary terms.  An alternative formula for the action is obtained by expanding $a$ around the background $a_0$.  The bulk integrand is
\be \CS(a_0 + b) = \CS(a_0) + 2\tr\big(b\wedge f(a_0)\big) - \dift\tr\big(a_0\wedge b\big) + \tr\bigg(b\wedge\Dift_0b + \frac{1}{3}b\wedge[b,b]\bigg) \ee
where $\Dift_0 = \dift + \ad_{a_0}$.  In the region $M_\eps$ the first and second terms vanish because $f(a_0) = 0$ and $a_{0,\bar w} = \p_{\bar w}a_{0,t} = \p_{\bar w}a_{0,x} = 0$.  In the action the third term passes to the boundary and cancels the source \eqref{eq:NLS-source}, so we're left with
\be \label{eq:NLS-4dCS+source}
S_\text{4dCS}[a;a_0] = S_\text{bulk}[a] + S_\text{source}[b;a_0] = \frac{\im}{2\pi}\int_M\frac{\dif w}{w^2}\wedge\tr\bigg(b\wedge\Dift_0b + \frac{1}{3}b\wedge[b,b]\bigg)\,.
\ee
The term involving the background is
\be \frac{\im}{2\pi}\int_M\frac{\dif w}{w^2}\wedge\tr(b\wedge[a_0,b]) = \frac{\im}{\pi}\int_M\dif t\wedge\dif x\wedge\frac{\dif w\wedge\dif \bar w}{w^4}\,\tr\big(b_{\bar w}[\im h,b_x + wb_t]\big)\,. \ee
The boundary conditions \eqref{eq:NLS-bcs}, \eqref{eq:NLS-wbar-bcs} give $[\im h,b_x + wb_t] = \mrm{O}(w^2)$, $b_{\bar w} = \mrm{O}(w)$ respectively, so the integrand is locally integrable.


\subsection{Determining the Lax} \label{subsec:NLS-Lax}

Let's determine the corresponding Lax with these boundary conditions.  To do so we implement a slight modification of the procedure introduced in \cite{Costello:2019tri}.  We begin by introducing a frame
\be a_{\bar w} = g^{-1}\p_{\bar w}g \ee
where $g:M\to\gSL_2(\bbC)$ and without loss of generality $g|_{w=0} = \mathrm{id}$.\footnote{Here we're assuming holomorphic triviality of the $\gSL_2(\bbC)$ bundle over $\CP^1_{(t,x)}$ determined by $\p_{\bar w} + a_{\bar w}$.  This is an open condition on the partial connection and holds in a neighbourhood of $b = 0$.}  Under a finite gauge transformation \eqref{eq:NLS-finite-gauge}, parameterised by $k:M\to\gSL_2(\bbC)$ and subject to the constraint $\log k\in\cI_\fh$ near $w=0$, the frame transforms by $g\mapsto gk^{-1}$.  We show in appendix \ref{app:gauge-fixing} that, subject to the boundary conditions $a_{\bar w}\in\cI_\fh$, the gauge invariant data is exhausted by
\be \label{eq:NLS-bundle-moduli} X = g^{-1}\p_wg|_{w=0} \ee
up to the redundancy $X\mapsto X - \p_w\log k|_{w=0}$.  Therefore $X$ determines an equivalence class $[X]\in \fsl_2(\bbC)/\fh$.  We will fix the representative
\be X = - \frac{\im}{2}(\psi e - \tilde\psi f) = - \frac{\im}{2}\begin{pmatrix} 0 & \psi \\ - \tilde\psi & 0 \end{pmatrix}\,. \ee
Next we impose the equations of motion with components in the $\bar w$ directions.  Writing
\be \label{eq:NLS-illegal-gauge} g^{-1}(\tilde\dif + a_0 + \Delta\cL)g = a_0 + b\,, \ee
where $\Delta\cL_{\bar w} = 0$, these imply that $\Delta\cL$ is meromorphic in $w$, potentially with poles at $w=0$.  $\Delta\cL$ is fixed by requiring that
\be \label{eq:NLS-gf-a} b = g^{-1}\tilde\dif g + g^{-1}a_0g - a_0 + g^{-1}\Delta\cL g\,. \ee
obeys the boundary conditions \eqref{eq:NLS-bcs} at $w=0$.

\begin{claim} \label{claim:NLS-Lax}
	The boundary conditions \eqref{eq:NLS-bcs} fix $\cL = a_0 + \Delta\cL$ to the Lax connection of the complexified NLS equation \eqref{eq:NLS-Lax} for the fields $\psi,\tilde\psi$.
\end{claim}

\begin{proof}
It's sufficient to determine $b_t$ to order $w^0$ and $b_x$ to order $w^1$.  In a neighbourhood of $w = 0$ we may take
\be g = \exp(wX)\,, \ee
though we emphasise that the terms of order $w^2$ and above are gauge dependent.  Let's now determine each of the three terms on the right-hand side of equation \eqref{eq:NLS-gf-a} in sequence.  The first can be computed using the derivative of the exponential map
\be \label{eq:NLS-Lax-pure-gauge} g^{-1}\dif_\Sigma g = - \frac{\im}{2}w(\dif_\Sigma\psi e - \dif_\Sigma\tilde\psi f) + \mrm{O}(w^2) \ee
Conjugation of the background gives
\bea \label{eq:NLS-background-conjugation}
&g^{-1}a_0g - a_0 = \bigg( - \frac{1}{w}\ad_X\im h + \frac{1}{2}\ad_X^2\im h + \mrm{O}(w)\bigg)\dif t \\
&+ \bigg(\ad_X\im h - \frac{1}{2}w\ad_X^2\im h + \mrm{O}(w^2)\bigg)\dif x
\eea
where 
\be \ad_X\im h = - \psi e - \tilde\psi f\,,\qquad \ad_X^2\im h = \im\psi\tilde\psi h\,. \ee
In order to evaluate the final term we'll need to know the pole structure of $\Delta\cL$, which is constrained by the requirement that $b$ remains finite as $w\to0$.  We've seen that \eqref{eq:NLS-Lax-pure-gauge} is free from singularities and \eqref{eq:NLS-background-conjugation} has a simple pole in the $t$ component, fixing
\be \Delta\cL = \bigg(A_t - \frac{1}{w}P\bigg)\dif t + A_x\dif x\,. \ee
Conjugating by $g^{-1}$ gives
\be g^{-1}\Delta\cL g = \bigg(A_t + \ad_XP - \frac{1}{w}P + \mrm{O}(w)\bigg)\dif t + \big(A_x - w\ad_XA_x + \mrm{O}(w^2)\big)\dif x\,. \ee
Having evaluated all three terms on the right-hand side of equation \eqref{eq:NLS-gf-a} we can impose the boundary conditions \eqref{eq:NLS-bcs}.
\begin{itemize}
	\renewcommand{\labelitemi}{$\blacktriangleright$}
	\item $b_t = \mrm{O}(1)$ as $w\to0$ implies
	\be P = - \ad_X\im h = e\psi + f\tilde\psi = \begin{pmatrix} 0 & \psi \\ \tilde\psi & 0 \end{pmatrix}\,. \ee
	\item $b_x = \mrm{O}(w)$ as $w\to0$ implies
	\be A_x = - \ad_X\im h = e\psi + f\tilde\psi\,. \ee
	\item $wb_t^e + b_x^e = \mrm{O}(w^2)$ and $wb_t^f + b_x^f = \mrm{O}(w^2)$ as $w\to0$ imply that
	\bea
	A_t^e = \frac{\im}{2}\p_x\psi\,,\qquad A_t^f = - \frac{\im}{2}\p_x\tilde\psi\,.
	\eea
	\item $b_t^h = \mrm{O}(w)$ as $w\to0$ implies
	\be A_t^h = - \frac{1}{2}(\ad_X^2\im h)^h - (\ad_XP)^h = \frac{\im}{2}\tilde\psi\psi\,. \ee
\end{itemize}
Putting this all together we have
\be \label{eq:NLS-Lie-Lax} \Delta\cL_t = - \frac{1}{w}(\psi e + \tilde\psi f) + \frac{\im}{2}\big(\p_x\psi e - \p_x\tilde\psi f + \tilde\psi\psi h\big)\,,\qquad \Delta\cL_x = e\psi + f\tilde\psi\,, \ee
or equivalently in matrix form
\be \cL_t + \frac{1}{w}\cL_x = \frac{\im}{2}\begin{pmatrix} \tilde\psi\psi & \p_x\psi \\ - \p_x\tilde\psi & - \tilde\psi\psi \end{pmatrix}\,,\qquad \cL_x = - \frac{1}{w}\begin{pmatrix} \im & 0 \\ 0 & -\im \end{pmatrix} + \begin{pmatrix} 0 & \psi \\ \tilde\psi & 0 \end{pmatrix}\,. \ee
We recognise this as the Lax for the complexified NLS equations \eqref{eq:NLS-Lax}.
\end{proof} 

The remaining unimposed equation of motion is equivalent to the zero-curvature condition on the Lax connection.  We can compute
\bea
&\p_t\cL_x = \begin{pmatrix} 0 & \p_t\psi \\ \p_t\tilde\psi & 0 \end{pmatrix}\,,\qquad \p_x\cL_t = - \frac{1}{w}\begin{pmatrix} 0 & \p_x\psi \\ \p_x\tilde\psi & 0 \end{pmatrix} + \frac{\im}{2}\begin{pmatrix} \p_x(\psi\tilde\psi) & \p_x^2\psi \\ - \p_x^2\tilde\psi & - \p_x(\psi\tilde\psi) \end{pmatrix}\,, \\
&[\cL_t,\cL_x] = - \frac{1}{w}\begin{pmatrix} 0 & \p_x\psi \\ \p_x\tilde\psi & 0 \end{pmatrix} + \frac{\im}{2}\begin{pmatrix} \p_x(\psi\tilde\psi) & 2\tilde\psi\psi^2 \\ - 2\tilde\psi^2\psi & -\p_x(\psi\tilde\psi) \end{pmatrix}\,.
\eea
The $w$ dependent terms cancel, as do those proportional to $h$, leaving
\be \p_t\cL_x - \p_x\cL_t + [\cL_t,\cL_x] = \begin{pmatrix} 0 & \p_t\psi - \frac{\im}{2}\p_x^2\psi + \im\tilde\psi\psi^2 \\ \p_t\tilde\psi + \frac{\im}{2}\p_x^2\tilde\psi - \im\tilde\psi^2\psi & 0 \end{pmatrix} = 0\,, \ee
which we recognise as the non-linear Schr\"{o}dinger equations \eqref{eq:NLS-equation-complex}
\be \label{eq:NLS-equations} \im\p_t\psi = - \frac{1}{2}\p_x^2\psi + \tilde\psi\psi^2\,,\qquad - \im\p_t\tilde\psi = - \frac{1}{2}\p_x^2\tilde\psi + \tilde\psi^2\psi\,. \ee


\subsection{Compactification} \label{subsec:NLS-compactify}

At this stage the partial connection is completely determined up to gauge in terms of $\psi,\tilde\psi$, and we can substitute in the regulated 4d Chern--Simons action and integrate over $\CP^1$ to recover a 2d functional.

\begin{claim} \label{claim:NLS-compactify}
	The regulated 4d Chern--Simons functional \eqref{eq:NLS-4dCS+source} restricted to the partially on-shell configuration \eqref{eq:NLS-gf-a} evaluates to the following action for the NLS equation
	\be S_{\bbC\mathrm{NLS}}[\psi,\tilde\psi] = \int_\Sigma\dif t\wedge\dif x\,\bigg( - \im\tilde\psi\p_t\psi + \frac{1}{2}\p_x\tilde\psi\p_x\psi + \frac{1}{2}\tilde\psi^2\psi^2\bigg)\,. \ee 
\end{claim}

\begin{proof}
	The 4d Chern--Simons action \eqref{eq:NLS-4dCS+source} can be written as the sum of a bulk and source term.  Let's begin by concentrating on the former.
	
	\paragraph{Bulk term.} Under a gauge transformation
	\be a = g^{-1}\dift g + g^{-1}\cL g \ee
	the mixed holomorphic-topological Chern--Simons three-form transforms by
	\be \label{eq:CS-gauge} \CS(a) = \CS(\cL) + \dift\tr\big(\rho\wedge\cL\big) + \frac{1}{3}\tr\big(\rho\wedge\rho\wedge\rho\big) \ee
	where $\rho = - \dift gg^{-1}$.  The first term does not contribute in the regulated action, since on $M_\eps$ we have $\cL_{\bar w} = 0 $ and $\p_{\bar w}\cL_t = \p_{\bar w}\cL_x = 0$.
	
	The final term of equation \eqref{eq:CS-gauge} also cannot contribute.  Indeed, we can fix a gauge representative
	\be g = \exp\big(wf(|w|)X\big) \ee
	for $f:[\eps,\infty)\to[0,1]$ a smooth bump function taking the value $1$ in an open neighbourhood of $\eps$ and $0$ for sufficiently large arguments.  Under phase rotations $w\mapsto e^{\im\theta}w$ the weight of $\tr(\rho\wedge\rho\wedge\rho)$ is at least $3$ whereas $\omega$ has weight $-1$.  The integral over the phase of $w$ therefore gives zero.
	
	The bulk 4d Chern--Simons action therefore localises to the boundary
	\be S_\text{bulk}[a] = - \int_\Sigma\Res_{w=0}\frac{\dif w}{w^2}\tr(\rho\wedge\cL) = \int_\Sigma\dif t\wedge\dif x\,\Res_{w=0}\frac{\dif w}{w^2}\tr(\rho_x\cL_t - \rho_t\cL_x)\,. \ee
	The components $\cL_t,\cL_x$ have at worst double and simple poles as $w\to0$, so we need only evaluate $\rho_x,\rho_t$ up to order $w^3,w^2$ respectively.  Using the derivative of the exponential map
	\be \rho = - w\dif_\Sigma X - \frac{w^2}{2}\ad_X\dif_\Sigma X - \frac{w^3}{6}\ad_X^2\dif_\Sigma X + \mrm{O}(w^4)\,. \ee
	Writing $X = - \frac{\im}{2}(e\psi - f\tilde\psi)$ we can evaluate
	\bea \label{eq:NLS-rhox-coefficients}
	&\p_xX = - \frac{\im}{2}(e\p_x\psi - f\p_x\tilde\psi)\,,\qquad \ad_X\p_xX = - \frac{h}{4}(\tilde\psi\p_x\psi - \psi\p_x\tilde\psi)\,, \\
	&\ad_X^2\p_xX = - \frac{\im}{4}(e\psi + f\tilde\psi)(\tilde\psi\p_x\psi - \psi\p_x\tilde\psi)\,,
	\eea
	and from equation \eqref{eq:NLS-Lie-Lax}
	\be \cL_t = \frac{1}{w^2}\im h - \frac{1}{w}(\psi e + f\tilde\psi) + \frac{\im}{2}\big(\p_x\psi e - \p_x\tilde\psi f + \psi\tilde\psi h\big) \ee
	yielding
	\be \Res_{w=0}\frac{\dif w}{w^2}\tr(\rho_x\cL_t) = \frac{1}{2}\p_x\tilde\psi\p_x\psi\,. \ee
	Similarly
	\be \label{eq:NLS-rhot-coefficients} \p_tX = - \frac{\im}{2}(e\p_t\psi - f\p_t\tilde\psi)\,,\qquad \ad_X\p_tX = - \frac{h}{4}(\tilde\psi\p_t\psi - \psi\p_t\tilde\psi) \ee
	and from equation \eqref{eq:NLS-Lie-Lax}
	\be \cL_x = - \frac{1}{w}\im h + e\psi + f\tilde\psi\,. \ee
	so that
	\be \Res_{w=0}\frac{\dif w}{w^2}\tr(\rho_t\cL_x) = \frac{\im}{4}(\tilde\psi\p_t\psi - \psi\p_t\tilde\psi)\,. \ee
	As a result the bulk part of the action evaluates to
	\be \label{eq:NLS-bulk-gf} S_\text{bulk}[a] = \int_\Sigma\dif t\wedge\dif x\,\bigg( - \frac{1}{2}\im\tilde\psi\p_t\psi + \frac{1}{2}\p_x\tilde\psi\p_x\psi\bigg)\,. \ee
	
	\paragraph{Source term.}  Next we turn our attention to the source term
	\be S_\text{source}[b] = - \int_\Sigma\Res_{w=0}\frac{\dif w}{w^2}\wedge\tr\big(a_0\wedge(g^{-1}\dif_\Sigma g + g^{-1}a_0g - a_0 + g^{-1}\Delta\cL g)\big)\,. \ee
	The trace can be simplified by noting that
	\be \tr(a_0\wedge g^{-1}a_0g) = \tr(a_0\wedge a_0) = 0\,, \ee
	simply because $a_0$ is proportional to $\dif t - w \dif x$.  The derivative of the exponential map gives
	\be g^{-1}\dif_\Sigma g = w\dif_\Sigma X - \frac{w^2}{2}\ad_X\dif_\Sigma X + \frac{w^3}{6}\ad_X^2\dif_\Sigma X + \mrm{O}(w^4)\,, \ee
	and we've evaluated the relevant coefficients in equations \eqref{eq:NLS-rhox-coefficients} and \eqref{eq:NLS-rhot-coefficients} so we get
	\be \label{eq:NLS-source-derivative} \Res_{w=0}\frac{\dif w}{w^4}\tr\big(\im h(g^{-1}\p_xg + wg^{-1}\p_tg)\big) = \frac{\im}{4}(\tilde\psi\p_t\psi - \psi\p_t\tilde\psi)\,. \ee
	The term involving the Lax is
	\bea
	&\Res_{w=0}\frac{\dif w}{w^2}\wedge\tr\big(a_0\wedge g^{-1}\Delta\cL g\big) = \dif t\wedge\dif x\,\Res_{w=0}\frac{\dif w}{w^4}\tr\big(\im hg^{-1}(\Delta\cL_x + w\Delta\cL_t)g\big) \\
	&= \dif t\wedge\dif x\,\Res_{w=0}\frac{\dif w}{w^3}\tr\big(\im hg^{-1}A_tg\big) = \dif t\wedge\dif x\,\frac{1}{2}\tr(\im h\ad_X^2A_t)\,,
	\eea
	which we can evaluate using
	\be \ad_X^2A_t = \frac{\im}{4}(\psi e + \tilde\psi f)(\tilde\psi\p_x\psi - \psi\p_x\tilde\psi) + \frac{\im}{2}h\tilde\psi^2\psi^2 \ee
	to get
	\be \label{eq:NLS-source-Lax} \Res_{w=0}\frac{\dif w}{w^3}\wedge\tr\big(\im hg^{-1}A_tg\big) = - \frac{1}{2}\tilde\psi^2\psi^2\,. \ee
	Collecting equations \eqref{eq:NLS-source-Lax} and \eqref{eq:NLS-source-derivative} we find
	\be \label{eq:NLS-source-gf} S_\text{source}[a] = \int_\Sigma\dif t\wedge\dif x\,\bigg(-\frac{\im}{2}\tilde\psi\p_t\psi + \frac{1}{2}\tilde\psi^2\psi^2\bigg)\,. \ee
	
	\paragraph{Collecting terms.} Finally putting together the bulk \eqref{eq:NLS-bulk-gf} and source terms \eqref{eq:NLS-source-gf} we recover the claimed action
	\be S_\text{4dCS}[a;a_0] 
	= \int_\Sigma\dif t\wedge\dif x\,\bigg( - \im\tilde\psi\p_t\psi + \frac{1}{2}\p_x\tilde\psi\p_x\psi + \frac{1}{2}\tilde\psi^2\psi^2\bigg) = S_{\bbC\text{NLS}}[\psi,\tilde\psi]\,. \ee
\end{proof}


\subsection{Real Structures} \label{subsec:NLS-reality}

In order to recover the attractive or repulsive NLS equation for a single complex valued field $\psi\in C^\infty(\Sigma;\bbC)$ we impose equivariance on the partial connection $a$ under simultaneous pullback by complex conjugation on the $\CP^1$ factor of $M$
\be \sigma:M\to M\,,\qquad (t,x,w,\bar w)\mapsto (t,x,\bar w,w) \ee
and an anti-holomorphic involution of the gauge algebra
\be \label{eq:NLS-conjugation} \tau:\fsl_2(\bbC)\mapsto \fsl_2(\bbC)\,,\qquad S \mapsto - \gamma^{-1}S^\dagger\gamma \ee
where $\gamma$ is a non-degenerate Hermitian form which can either be positive definite or split.  Working in the basis of the fundamental in which $\im h$ is diagonal we take $\gamma = \delta = \diag(1,1)$ in the positive definite case or $\gamma = \eta = \diag(1,-1)$ in the split case.  The fixed point subalgebra of $\tau$ is $\fsu(2)$ or $\fsu(1,1)$ respectively.  Gauge transformations are similarly required to be invariant under simultaneous pullback by $\sigma$ and the lift of $\tau$ to $\gSL_2(\bbC)$.  The background $a_0$ and boundary conditions on $b$ are invariant under \eqref{eq:NLS-conjugation}.  

These real structures imply that $X$ from equation \eqref{eq:NLS-bundle-moduli} is valued in the fixed point subalgebra.  In the case of a positive definite Hermitian form this yields
\be - X^\dagger = - \frac{\im}{2}\begin{pmatrix} 0 & - \bar{\tilde\psi} \\ \bar\psi & 0 \end{pmatrix} = - \frac{\im}{2}\begin{pmatrix} 0 & \psi \\ - \tilde\psi & 0 \end{pmatrix} = X\,, \ee
so that $\tilde\psi = -\bar\psi$ recovering the attractive NLS equation.  $S_{\bbC\text{NLS}}[\psi,-\bar\psi]$ coincides with the real action \eqref{eq:NLS-action} on the nose.

In the split case we have
\be - \eta^{-1}X^\dagger\eta = - \frac{\im}{2}\begin{pmatrix} 0 & \bar{\tilde\psi} \\ - \bar\psi & 0 \end{pmatrix} = - \frac{\im}{2}\begin{pmatrix} 0 & \psi \\ - \tilde\psi & 0 \end{pmatrix} = X \ee
so that $\tilde\psi = \bar\psi$, leading to the repulsive NLS equation.  $S_{\bbC\text{NLS}}[\psi,\bar\psi]$ differs from the real action \eqref{eq:NLS-action} by an overall sign.


\section{Korteweg--de Vries} \label{sec:KdV}

Next let's turn our attention to the KdV equation
\be \label{eq:KdV-equation} \p_tu - \frac{1}{4}\p_x^3u - 3u\p_xu = 0\,, \ee
which describes shallow water waves in one space and one time dimension.  In order to write down a Lax and variational principle for the KdV equation we write $u = \p_xv$ for a potential $v\in C^\infty(\Sigma,\bbR)$ defined modulo
\be \label{eq:KdV-shift-symmetry} v\mapsto v + \beta(t)\,. \ee
A Lax connection for the KdV equation is given by
\be \label{eq:KdV-Lax} \cL_t - z\cL_x = \begin{pmatrix} \frac{1}{2}\p_x^2v + v\p_xv & -\p_xv \\ c & - \frac{1}{2}\p_x^2v - v\p_xv \end{pmatrix}\,,\qquad \cL_x = \begin{pmatrix} v & -1 \\ \p_xv + v^2 - z & - v \end{pmatrix} \ee
for $c\in C^\infty(\Sigma,\bbR)$ an independent field.  The zero-curvature equation fixes
\be c = - \p_tv + \frac{1}{2}\p_x^3v + 2(\p_xv)^2 + v\p_x^2v + v^2\p_xv\,, \ee
but we'll recover the Lax with $c$ undetermined.  A variational formulation is supplied by
\be \label{eq:S-KdV} S_\mathrm{KdV}[v] = 2\int_\Sigma\dif t\wedge\dif x\,\bigg(\p_tv\p_xv + \frac{1}{4}(\p_x^2v)^2 - (\p_xv)^3\bigg)\,. \ee

In this section we'll see that the KdV equation can be obtained as a compactification of 4d Chern--Simons theory.  We begin in section \ref{subsec:KdV-bcs} by introducing the singular boundary conditions for 4d Chern--Simons which lead to KdV.  These closely resemble those used to recover non-linear Schr\"{o}dinger; the crucial difference is that the coefficient of the double pole in the background field $a_0$ is taken to be the nilpotent element $-f\in\fsl_2(\bbC)$ rather than Cartan valued.  The use of singular boundary conditions requires the introduction of source terms localised at $w=0$, which are determined in section \ref{subsec:KdV-source}.  We show in section \ref{subsec:KdV-Lax} that, upon restriction to the $\CP^1_{(t,x)}$ fibre over a point in $\Sigma$, the partial connection is determined by two complex parameters which we identify with the potential $v$ and an independent auxiliary field $r$.  The redundancy \eqref{eq:KdV-shift-symmetry} also has a direct interpretation as a gauge symmetry with restricted co-ordinate dependence.  We further demonstrate that working partially on the support of the equations of motion the partial connection $a$ is gauge equivalent to the Lax of the KdV equation \eqref{eq:KdV-Lax}.  In the process the auxiliary field is fixed to $r = \tfrac{1}{2}(\p_xv + v^2)$, although a new independent field $c$ emerges.  In section \ref{subsec:KdV-compactify} we evaluate the 4d Chern--Simons action on this configuration.  The field $c$ drops out and we recover the functional \eqref{eq:S-KdV}.  Throughout we largely work with complexified fields, though in section \ref{subsec:KdV-reality} we explain how to recover a real potential $v$.


\subsection{Boundary Conditions} \label{subsec:KdV-bcs}

As in the case of NLS, we take $G = \gSL_2(\bbC)$, $C=\CP^1$ and the meromorphic one-form to be
\be \omega = \dif z = \frac{\dif w}{w^2}\,. \ee
Recall that the pre-symplectic structure specialises to
\bea \label{eq:KdV-symplectic}
&\Omega = 2\int_\Sigma\Res_{w=0}\frac{\dif w}{w^2}\wedge\dif t\wedge\dif x\wedge\tr(\delta a_t\wedge\delta a_x) \\
&= 2\int_\Sigma\dif t\wedge\dif x\wedge\Res_{w=0}\frac{\dif w}{w^2}\wedge\big(\delta a_t^e\wedge\delta a_x^f + \delta a_t^f\wedge\delta a_x^e + 2\delta a_t^h\wedge\delta a_x^h\big)\,.
\eea
Our boundary conditions on $a$ will need to be isotropic for this form.

We prescribe fixed singular behaviour on the space-time components of the gauge field as $w\to0$, and impose restrictions on fluctuations around this background.  That is, we fix
\be \label{eq:KdV-split} a = a_0 + b \ee
where
\be \label{eq:KdV-a0} a_0 = - (zf + e)(z\dif t + \dif x) = - \frac{1}{w^2}(f - we)(\dif t - w\dif x) \ee
solves the 4d Chern--Simons equations of motion away from $\Sigma_0$ and $b$ is non-singular.  We further require that the space-time components of $b$ obey the boundary conditions
\bea \label{eq:KdV-bcs}
&b_t^e = \mrm{O}(w)\,,\qquad b_t^f = \mrm{O}(1)\,,\qquad wb_t^h + b_x^h = \mrm{O}(w^2)\,, \\
&b_x^e = \mrm{O}(w^2)\,,\qquad b_x^f = \mrm{O}(w)\,,\qquad b_x^h = \mrm{O}(w)\quad\text{as}\quad w\to0\,.
\eea
Since $a_0$ is fixed the variations of $a$ and $b$ coincide: $\delta a = \delta b$.  We can check the Lagrangian condition by determining for which $\xi\in\til\Omega^{0,1}(M;\fsl_2(\bbC))$ the pairing
\bea
&\Omega(\xi,b) = 2\int_\Sigma\dif t\wedge\dif x\Res_{w=0}\frac{\dif w}{w^2}\big(\xi_t^eb_x^f + \xi_t^fb_x^e + 2\xi_t^h(wb_t^h + b_x^h) \\
&- b_t^e\xi_x^f - b_t^f\xi_x^e - 2b_t^h(w\xi_t^h + \xi_x^h)\big)
\eea
is zero for all allowed $b$.  For the residue to vanish the term in brackets must be $\mrm{O}(w^2)$ as $w\to0$.  Hence
\bea
&\xi_t^e = \mrm{O}(w)\,,\qquad\xi_t^f = \mrm{O}(1)\,,\qquad\xi_t^h = \mrm{O}(1)\,, \\
&\xi_x^e = \mrm{O}(w^2)\,,\qquad\xi_x^f = \mrm{O}(w)\,,\qquad w\xi_t^h + \xi_x^h = \mrm{O}(w^2)\quad\text{as}\quad w\to0\,.
\eea
Since $\xi_t^h$ is non-singular then the final condition implies that $\xi_x^h = \mrm{O}(w)$.  The annihilator of the subspace cut out by the boundary conditions \eqref{eq:KdV-bcs} is itself, so it is Lagrangian.

Whilst these boundary conditions are isotropic for the pre-symplectic structure, they do not annihilate the pre-symplectic potential.  We are therefore required to introduce compensating source terms supported at $w=0$ which are determined in section \ref{subsec:KdV-source}.

Recall that an infinitesimal gauge transformation with parameter $\eps\in C^\infty(M;\fsl_2(\bbC))$ acts by
\be \delta_\eps a = \delta_\eps b = \dift\eps + [a_0,\eps] + [b,\eps]\,. \ee
We must restrict $\eps$ so that the boundary conditions \eqref{eq:KdV-bcs} are preserved.

\begin{claim}
	Infinitesimal gauge transformations $\eps = \eps^ee + \eps^ff + \eps^hh$ obeying
	\be \label{eq:KdV-eps-bcs} \eps^e = \mrm{O}(w^2)\,,\qquad\eps^f = \mrm{O}(w)\,,\qquad \eps^h = \mrm{O}(w^2)\,,\quad\text{as}\quad w\to0 \ee
	respect the boundary conditions \eqref{eq:KdV-bcs}.
\end{claim}

\begin{proof}
Since $\eps = \mrm{O}(w)$ and $b$ is non-singular the terms $\dift\eps$ and $[b,\eps]$ respect all but $\delta a^e_x = \mrm{O}(w^2)$ and $w\delta a_t^h + \delta a_x^h = \mrm{O}(w^2)$.  The stronger constraints $\eps^e = \eps^h = \mrm{O}(w^2)$ ensure that the $\dift\eps$ term also respects these conditions.  Furthermore
\bea
&[b,\eps]^e = 2b^h\eps^e - 2b^e\eps^h = \mrm{O}(w^2)\,, \\
&[b,\eps]^h = b^e\eps^f - b^f\eps^e = \mrm{O}(w^2)\,.
\eea
Here the first line holds since $\eps^e = \eps^h = \mrm{O}(w^2)$.  The second holds despite the fact that $\eps^f = \mrm{O}(w)$ because $b^e = \mrm{O}(w)$.

This leaves the background term.  Let's write $\eps = wFf + w^2S$ for $F\in C^\infty(M)$ and $S\in C^\infty(M;\fsl_2(\bbC))$, so that
\bea
&[a_{0,t},\eps] = Fh - [f,S] + w[e,S] = (F + S^e + wS^f)h - 2S^hf - 2wS^he\,, \\
&[a_{0,x},\eps] = - wFh + w[f,S] - w^2[e,S] = - w(F + S^e + wS^f)h + 2wS^hf + 2w^2S^he\,.
\eea
Comparing with the conditions \eqref{eq:KdV-bcs} we can see that all are obeyed.
\end{proof}

More abstractly, we can define the Lie subalgebra $\cI_{\fn_-}$ of $C^\infty(M;\fg)$ for $\fn_- = \la f\ra_\bbC$.  $\cI_{\fn_-}$ consists of all smooth maps $\eta\in C^\infty(M;\fg)$ obeying
\be \eta^e = \mrm{O}(w^2)\,,\qquad \eta^f = \mrm{O}(w)\,,\qquad \eta^h = \mrm{O}(w^2)\,,\quad\text{as}\quad w\to0\,. \ee
It obeys $[\cI_{\fn_-},\cI_{\fn_-}]\subset\cI_{\fn_-}$.  Then our boundary conditions on infinitesimal gauge transformations state that $\eps\in\cI_{\fn_-}$.  The finite counterpart of an infinitesimal gauge transformation is a group valued map $k:M\to\gSL_2(\bbC)$ acting on the partial connection $a$ by
\be \label{eq:KdV-finite-gauge} a_0 + b \mapsto k\dift k^{-1} + k(a_0 + b)k^{-1}\,. \ee
The boundary conditions on infinitesimal gauge transformations \eqref{eq:KdV-eps-bcs} translate into the condition $\log k\in\cI_{\fn_-}$ in a neighbourhood of $w=0$.  (Note that $\log$ has a unique branch in a sufficiently small neighbourhood of the identity.)

Finally we need to impose boundary conditions on the $\dif\bar w$ components of $a$.  We will simply require that $a_{\bar w}\in\cI_{\fn_-}$, i.e.,
\be \label{eq:KdV-wbar-bcs} a_{\bar w}^e = \mrm{O}(w^2)\,,\qquad a_{\bar w}^f = \mrm{O}(w)\,,\qquad a_{\bar w}^h = \mrm{O}(w^2)\quad\text{as}\quad w\to0 \ee
which is a consistent boundary condition since $\p_{\bar w}\cI_{\fn_-}\subset\cI_{\fn_-}$ and $[\cI_{\fn_-},\cI_{\fn_-}]\subset\cI_{\fn_-}$.  As in section \ref{subsec:NLS-bcs}, there is no co-isotropy condition on $a_{\bar w}$ since it does not enter the pre-symplectic form.

4d Chern--Simons with the above boundary conditions inherits a symmetry, somewhat analogous to the current algebra symmetries of the 2d Wess-Zumino-Witten model, generated by gauge transformations which obey slightly weaker boundary conditions than equations \eqref{eq:KdV-eps-bcs} but with restricted dependence on the space-time co-ordinates.  In particular infinitesimal gauge transformations obeying
\be \label{eq:KdV-constrained-gauge} \eps^e = w\beta(t) + \mrm{O}(w^2)\,,\qquad \eps^f = - \beta(t) + \mrm{O}(w)\,,\qquad \eps^h = \mrm{O}(w^2)\,. \ee
for $\beta(t)$ a smooth function of $t$ also respect the boundary conditions \eqref{eq:KdV-bcs}.  Whether one treats these as true gauge transformations is a choice, but they represent a real ambiguity in the evolution equations.

Away from $\Sigma_0 = \{w=0\}\subset M$ the background connection $a_0$ obeys the 4d Chern--Simons equations of motion, but on this locus it's sourced.  We now determine the form of these source terms.


\subsection{Source Terms} \label{subsec:KdV-source}

We can determine the source terms which must be adjoined to the action by finding a potential for the pre-symplectic potential \eqref{eq:symplectic-potential} when restricted to the boundary conditions of the previous section.  Recall that the pre-symplectic potential is
\be
\Theta = - \int_\Sigma\Res_{w=0}\frac{\dif w}{w^2}\tr(a\wedge \delta a) = \int_\Sigma\dif t\wedge\dif x\,\Res_{w=0}\frac{\dif w}{w^2}\tr\big(a_t\delta a_x - a_x\delta a_t\big)\,,
\ee
which for our pinned boundary conditions restricts to
\bea
&\Theta|_\text{bc} = - \int_\Sigma\Res_{w=0}\frac{\dif w}{w^2}\tr(a_0\wedge \delta a) \\
&= - \int_\Sigma\dif t\wedge\dif x\,\mrm{Res}_{w=0}\frac{\dif w}{w^4}\tr\big((f-ew)(\delta a_x + w\delta a_t)\big)\,.
\eea
This can be compensated by the source term
\bea \label{eq:KdV-source}
&S_\text{source}[b;a_0] = - \int_\Sigma\mrm{Res}_{w=0}\frac{\dif w}{w^2}\wedge\tr(a_0\wedge b) \\
&= \int_\Sigma\dif t\wedge\dif x\,\mrm{Res}_{w=0}\frac{\dif w}{w^4}\tr\big((f-ew)(b_x + wb_t)\big)\,.
\eea
We should check that the source term is gauge invariant.

\begin{claim}
	The source term \eqref{eq:KdV-source} is invariant under infinitesimal gauge transformations of the form \eqref{eq:KdV-eps-bcs}.
\end{claim}
 
\begin{proof}
Under an infinitesimal gauge transformation
\be
\delta_\eps S_\text{source}[b;a_0] = - \int_\Sigma\mrm{Res}_{w=0}\frac{\dif w}{w^2}\wedge\tr\big(a_0\wedge\dif_\Sigma\eps + a_0\wedge[a_0,\eps] + a_0\wedge[b,\eps]\big)\,. \ee
The first term in the trace is exact on $\Sigma$ and the second vanishes because $[a_0,a_0] = 0$.  This leaves
\bea \label{eq:gauge-Sbdry}
&\delta_\eps S_\text{source}[b;a_0] = - \int_\Sigma\mrm{Res}_{w=0}\frac{\dif w}{w^2}\wedge\tr\big(a_0\wedge[b,\eps]\big) \\
&= \int_\Sigma\dif t\wedge\dif x\Res_{w=0}\frac{\dif w}{w^4}\wedge\big([b_x + wb_t,\eps]^e - w[b_x + wb_t,\eps]^f\big)
\eea
Writing out the commutators 
\bea
&[b_x + wb_t,\eps]^e = 2(b_x^h + wb_t^h)\eps^e - 2(b_x^e + wb_t^e)\eps^h\,, \\
&[b_x + wb_t,\eps]^f = 2(b_x^f + wb_t^f)\eps^h - 2(b_x^h + wb_t^h)\eps^f\,.
\eea
The boundary conditions \eqref{eq:KdV-bcs} imply that the first line is $\mrm{O}(w^4)$, for example, $b_x^h + wb_t^h = \mrm{O}(w^2)$ and $\eps^e = \mrm{O}(w^2)$.  Similarly the second line is $\mrm{O}(w^3)$, so the gauge variation of the boundary term \eqref{eq:gauge-Sbdry} vanishes.
\end{proof}

Expanding $a = a_0 + b$ we can rewrite the full bulk and boundary action as
\be \label{eq:KdV-4dCS+source}
S_\text{4dCS}[a;a_0] = S_\text{bulk}[a_0 + b] + S_\text{source}[b;a_0] = \frac{\im}{2\pi}\int_M\frac{\dif w}{w^2}\wedge\tr\bigg(b\wedge\Dift_0b + \frac{1}{3}b\wedge [b,b]\bigg)
\ee
where $\Dift_0 = \dift + \ad_{a_0}$.  The term involving the background is
\be \frac{\im}{2\pi}\int_M\frac{\dif w}{w^2}\wedge\tr(b\wedge[a_0,b]) = - \frac{\im}{\pi}\int_M\dif t\wedge\dif x\wedge\frac{\dif w\wedge\dif \bar w}{w^4}\,\tr\big(b_{\bar w}[f - ew,b_x + wb_t]\big)\,. \ee
The boundary conditions \eqref{eq:KdV-bcs}, \eqref{eq:KdV-wbar-bcs} give $[f-ew,b_x + wb_t] = \mrm{O}(w^2)$, $b_{\bar w} = \mrm{O}(w)$ respectively, so the integrand is locally integrable.


\subsection{Determining the Lax} \label{subsec:KdV-Lax}

To determine the Lax we begin by fixing a frame
\be a_{\bar w} = g^{-1}\p_{\bar w}g \ee
for $g:M\to\gSL_2(\bbC)$ where $g|_{w=0} = \mathrm{id}$.  Under a finite gauge transformation $k:M\to\gSL_2(\bbC)$ obeying $k|_{w=0} = \mathrm{id}$ and $k^{-1}\p_wk\in\cI_{\fn_-}$ the frame transforms by $g\mapsto gk^{-1}$.  The argument in appendix \ref{app:gauge-fixing} shows that the gauge invariant data is exhausted by 
\be B = g^{-1}\p_wg|_{w=0} \ee
up to the redundancy $B\mapsto B - \p_w\log k|_{w=0}$, so it determines an equivalence class $[B]\in\fsl_2(\bbC)/\mathfrak{n}_-$.  We can fix a representative $B\in\mathfrak{b}^+$ parametrised by
\be \label{eq:KdV-representative} B = ve - rh = \begin{pmatrix} - r & v \\ 0 & r \end{pmatrix} \ee
where $v,r\in C^\infty(\Sigma;\bbC)$ are at this stage independent fields.

Next we impose the equations of motion with components in the $\bar w$ directions, which imply that $\Delta\cL$ is meromorphic in $w$, potentially with poles at $w = 0$, where
\be \label{eq:KdV-illegal-gauge} g^{-1}(\dift + a_0 + \Delta\cL)g = a_0 + b\,. \ee
Equivalently
\be \label{eq:KdV-gf-a} b = g^{-1}\dift g + g^{-1}a_0g - a_0 + g^{-1}\Delta\cL g\,. \ee
$\Delta\cL$ is fixed by requiring that $b$ obeys the boundary conditions \eqref{eq:KdV-bcs} at $w=0$.

\begin{claim} \label{claim:KdV-Lax}
	The boundary conditions \eqref{eq:KdV-bcs} fix $\cL = a_0 + \Delta\cL$ to the Lax connection of the KdV equation \eqref{eq:KdV-Lax} for $u = \p_xv$.
\end{claim}

\begin{proof}
It's sufficient to determine $b_t$ to order $w^0$ and $b_x$ to order $w^1$.  In a neighbourhood of $w = 0$ we may take
\be g = \exp(wB)\,, \ee
We now evaluate each of the three terms on the right-hand side of equation \eqref{eq:KdV-gf-a} in this neighbourhood.  The first can be computed using the derivative of the exponential map
\be g^{-1}\dif_\Sigma g = w\dif_\Sigma B + \mrm{O}(w^2) = w\dif_\Sigma ve - w\dif_\Sigma rh + \mrm{O}(w^2)\,.
\ee
Conjugating the background field gives
\bea \label{eq:KdV-conj-a0}
&g^{-1}a_0g - a_0 = \bigg(\frac{1}{w}\ad_Bf - \ad_Be - \frac{1}{2}\ad_B^2f + \mrm{O}(w)\bigg)\dif t \\
&+ \bigg( - \ad_Bf + w\ad_Be + \frac{1}{2}w\ad_B^2f + \mrm{O}(w^2)\bigg)\dif x\,.
\eea
where
\be \ad_Bf = 2rf + vh\,,\qquad \ad_Be = - 2re\,,\qquad \ad_B^2f = 4r^2f + 2rvh - 2v^2e\,. \ee
Evaluating the final term in equation \eqref{eq:KdV-gf-a} requires knowing the pole structure of $\Delta\cL$, which can be fixed by finiteness of $b$ as $w\to0$.   Since \eqref{eq:KdV-conj-a0} has a simple pole in its $\dif t$ component at $w=0$, we must take
\be \Delta\cL = \bigg(A_t - \frac{1}{w}\Delta P\bigg)\dif t + \Delta A_x\dif x\,. \ee
Conjugating by $g^{-1}$ then gives
\be g^{-1}\Delta\cL g = \bigg(A_t + \ad_B\Delta P - \frac{1}{w}\Delta P + \mrm{O}(w)\bigg)\dif t + \big(\Delta A_x - w\ad_B\Delta A_x + \mrm{O}(w^2)\big)\dif x\,. \ee
Having evaluated all terms on the right-hand side of equation \eqref{eq:KdV-gf-a} we can now impose the boundary conditions \eqref{eq:KdV-bcs}:
\begin{itemize}
	\renewcommand{\labelitemi}{$\blacktriangleright$}
	\item $b_x = \mrm{O}(w)$ as $w\to0$ implies
	\be \Delta A_x = \ad_Bf =  2rf + vh = \begin{pmatrix} v & 0 \\ 2r & -v \end{pmatrix}\,. \ee
	\item $b_t = \mrm{O}(1)$ as $w\to0$ implies
	\be \Delta P = \ad_Bf = 2rf + vh\,. \ee
	\item $b_x^e = \mrm{O}(w^2)$ as $w\to0$ implies
	\be \label{eq:KdV-fix-r} \p_xB^e + (\ad_Be)^e + \frac{1}{2}(\ad_B^2f)^e - [B,\Delta A_x]^e = \p_xv + v^2 - 2r = 0\,. \ee
	So that $r = \tfrac{1}{2}(\p_xv + v^2)$.  This constraint is somewhat unusual since it imposes a relation on the moduli $(r,v)$.
	\item $b_t^e = \mrm{O}(w)$ as $w\to0$ implies
	\be A_t^e = (\ad_Be)^e + \frac{1}{2}(\ad_B^2f)^e - (\ad_B\Delta P)^e = v^2-2r = - \p_xv\,. \ee
	\item $wb_t^h + b_x^h = \mrm{O}(w^2)$ as $w\to0$ implies
	\be A_t^h + (\ad_B\Delta P)^h - (\ad_B\Delta A_x)^h + (\p_xB)^h = A_t^h - \p_xr = 0\,, \ee
	hence $A_t^h = \p_xr = \tfrac{1}{2}\p_x^2v + v\p_xv$.
\end{itemize}
Another unusual feature of this reduction is that the component $c\coloneqq A_t^f\in C^\infty(\Sigma;\bbC)$ remains unfixed.  Assembling all of these components together we have
\be \label{eq:KdV-Lie-Lax} \Delta\cL_t = - \frac{1}{w}(2rf + vh) + \big( - \p_xve + \p_xrh + cf\big)\,,\qquad \Delta\cL_x = 2rf + vh \ee
where $r = \tfrac{1}{2}(\p_xv + v^2)$.  In matrix form
\be \label{eq:KdV-matrix-Lax} \cL_t + \frac{1}{w}\cL_x = \begin{pmatrix} \frac{1}{2}\p_x^2v + v\p_xv & -\p_xv \\ c & - \frac{1}{2}\p_x^2v - v\p_xv \end{pmatrix}\,,\qquad \cL_x = \frac{1}{w}\begin{pmatrix} 0 & 0 \\ 1 & 0 \end{pmatrix} + \begin{pmatrix} v & -1 \\ \p_xv + v^2 & - v \end{pmatrix}\,. \ee
We recognise this as the Lax connection for the KdV equation \eqref{eq:KdV-Lax}.
\end{proof}

The remaining 4d Chern--Simons equation of motion is equivalent to the zero-curvature equation on the Lax \eqref{eq:KdV-matrix-Lax}.  Direct computation gives
\bea
&\p_t\cL_x = \begin{pmatrix} \p_tv & 0 \\ 2\p_tr & - \p_tv \end{pmatrix}\,,\qquad \p_x\cL_t = - \frac{1}{w}\begin{pmatrix} \p_xv & 0 \\ 2\p_xr & - \p_xv \end{pmatrix} + \begin{pmatrix} \p_x^2r & -\p_x^2v \\ \p_xc & - \p_x^2r \end{pmatrix} \\
&[\cL_t,\cL_x] = - \frac{1}{w}\begin{pmatrix} \p_xv & 0 \\ 2\p_xr & - \p_xv \end{pmatrix} + \begin{pmatrix} c - 2r\p_xv & - \p_x^2v \\ 2cv - 2\p_x(r^2) & 2r\p_xv - c \end{pmatrix}\,.
\eea
Putting this together
\be \p_t\cL_x - \p_x\cL_t + [\cL_t,\cL_x] = \begin{pmatrix} \p_tv - \p_x^2r - 2r\p_xv + c & 0 \\ 2\p_tr - \p_xc + 2cv - 2\p_x(r^2) & - \p_tv + \p_x^2r + 2r\p_xv - c \end{pmatrix} = 0\,. \ee
The diagonal terms fix $c$ in terms of $v$
\be c = - \p_tv + \p_x^2r + 2r\p_xv = - \p_tv + \frac{1}{2}\p_x^3v + 2(\p_xv)^2 + v\p_x^2v + v^2\p_xv\,, \ee
where in the second equality we've used $r = \tfrac{1}{2}(\p_xv + v^2)$.  The final equation is
\be 2\p_tr - \p_xc + 2cv - 2\p_x(r^2) = 0\,, \ee
and substituting in the known expressions for $c,r$ this becomes the KdV equation for $u = \p_xv$
\be \p_tu - \frac{1}{4}\p_x^3u - 3u\p_xu = 0\,. \ee

We should also understand the origin of the shift symmetry $v\mapsto v + \beta(t)$.  Under the constrained infinitesimal gauge transformation \eqref{eq:KdV-constrained-gauge}
\be \delta_\beta(g^{-1}\p_wg)|_{w=0} = \p_w\eps|_{w=0} + [g^{-1}\p_wg,\eps]|_{w=0}\,. \ee
To find the corresponding action on $B$ we should interpret this as an equation in $\fsl_2(\bbC)/\mf{n}_-$
\be \delta_\beta B \sim \begin{pmatrix} - \beta v & \beta \\ 2r\beta & \beta v \end{pmatrix}\,. \ee
For our representative this implies that $\delta_\beta v = \beta(t)$ and $\delta_\beta r = \beta(t)v$, which is consistent with the relation $r = \tfrac{1}{2}(\p_xv + v^2)$.


\subsection{Compactification} \label{subsec:KdV-compactify}

At this stage the $z$ dependence of $b$ is completely determined by \eqref{eq:KdV-gf-a}, allowing us to substitute it back in to the regulated 4d Chern--Simons action \eqref{eq:KdV-4dCS+source} and compactify to $\Sigma$.

\begin{claim} \label{claim:KdV-compactify}
	The regulated 4d Chern--Simons functional \eqref{eq:KdV-4dCS+source} restricted to the partially on-shell configuration \eqref{eq:KdV-gf-a} evaluates to the following action for the KdV equation
	\be S_\mathrm{KdV}[v] = 2\int_\Sigma\dif t\wedge\dif x\,\bigg(\p_tv\p_xv + \frac{1}{4}(\p_x^2v)^2 - (\p_xv)^3\bigg)\,. \ee 
\end{claim}

\begin{proof}
	The 4d Chern--Simons action \eqref{eq:KdV-4dCS+source} consists of a bulk and source piece which we'll evaluate separately, starting with the former.  Much of the general analysis follows the argument of section \ref{subsec:NLS-compactify}, and we will sometimes refer back to that argument when convenient. 
	
	\paragraph{Bulk term.}  The bulk piece is
	\be \label{eq:KdV-bulk-action} S_\text{bulk}[a] = \frac{\im}{2\pi}\int_M\frac{\dif w}{w^2}\wedge\CS(a)\,. \ee
	From equation \eqref{eq:KdV-illegal-gauge}, the partial connection and Lax are related by the boundary condition violating gauge transformation $a = g^{-1}(\dift + \cL)g$.  Under a gauge transformation the mixed holomorphic-topological Chern--Simons three-form transforms by
	\be \label{eq:KdV-CS-gauge} \CS(a) = \CS(\cL) + \dift\tr\big(\rho\wedge\cL\big) + \frac{1}{3}\tr\big(\rho\wedge\rho\wedge\rho\big) \ee
	where $\rho = - \dift gg^{-1}$.  As we saw in section \ref{subsec:NLS-compactify}, neither the first nor last term can contribute in the action \eqref{eq:KdV-bulk-action}, so the bulk 4d Chern--Simons action contributes only a boundary term	
	\be S_\text{bulk}[a] 
	= \int_\Sigma\dif t \wedge\dif x\,\mrm{Res}\frac{\dif w}{w^2}\tr\big(\rho_x\cL_t - \rho_t\cL_x\big)\,. \ee
	The residue picks out terms of order $w$.  We begin by concentrating on the first term in the trace.  Since $\cL_t$ has at worst a double pole in $w$, we need only expand $\rho_x$ to order $O(w^3)$.  Differentiating the matrix exponential gives
	\bea
	&\rho_x = - w\p_xB - \frac{w^2}{2}\ad_B\p_xB - \frac{w^3}{6}\ad_B^2\p_xB + \mrm{O}(w^4) \\
	&= w(\p_xr h - \p_xv e) + w^2(r\p_xv - v\p_xr)e + \frac{2}{3}w^3r(v\p_xr - r\p_xv)e + \mrm{O}(w^4)\,.
	\eea
	Recalling the form of the Lax \eqref{eq:KdV-Lie-Lax}
	\be \cL_t = - \frac{1}{w^2}f - \frac{1}{w}(- e + vh + 2rf)  -\p_xve + \p_xrh + cf \ee
	we can directly compute
	\be \Res_{w=0}\frac{\dif w}{w^2}\tr\big(\rho_x\cL_t\big) 
	= 2(\p_xr)^2 - c\p_xv + \frac{4}{3}vr\p_xr - \frac{4}{3}r^2\p_xv\,. \ee
	Repeating these steps for the second term in the trace
	\bea
	&\rho_t = - w\p_tB - \frac{w^2}{2}\ad_B\p_tB + \mrm{O}(w^3) = w(\p_tr h - \p_tv e) + w^2(r\p_tv - v\p_tr)e + \mrm{O}(w^3)\,, \\
	&\cL_x = \frac{1}{w}f - e + vh + 2rf
	\eea
	we find
	\be \op{Res}\frac{\dif w}{w^2}\tr\big(\rho_t\cL_x) = v\p_tr - r\p_tv\,. \ee
	The bulk part of the action therefore evaluates to
	\be \label{eq:KdV-bulk-evaluated} S_\text{bulk}[a] = \int_\Sigma\dif t \wedge\dif x\,\big(2r\p_tv + 2(\p_xr)^2 - 2r^2\p_xv - c\p_xv\big)\,. \ee
	
	\paragraph{Source term.} Next let us turn our attention to the source term
	\bea
	&S_\text{source}[a;a_0] = - \int_\Sigma\Res_{w=0}\frac{\dif w}{w^2}\wedge\tr(a_0\wedge b) \\ &
	= - \int_\Sigma\Res_{w=0}\frac{\dif w}{w^2}\wedge\tr\big(a_0\wedge(g^{-1}\dif_\Sigma g 
	+ g^{-1}\Delta\cL g)\big)\,.
	\eea
	Where we've discarded terms involving conjugation of the background field as in section \ref{subsec:NLS-compactify}.

	Differentiating the matrix exponential
	\bea
	&g^{-1}\dif_\Sigma g = w\dif_\Sigma B - \frac{w^2}{2}\ad_B\dif_\Sigma B + \frac{w^3}{6}\ad_B^2\dif_\Sigma B + \mrm{O}(w^4) \\
	&= w(\dif_\Sigma v e - \dif_\Sigma r h) + w^2(r\dif_\Sigma v - v\dif_\Sigma r)e + \frac{2}{3}w^3r(r\dif_\Sigma v - v\dif_\Sigma r)e + \mrm{O}(w^4)
	\eea
	Since this is upper triangular, it can only pair with $f$ component of $a_0$, yielding
	\bea
	&\Res_{w=0}\frac{\dif w}{w^2}\wedge\tr\big(a_0\wedge g^{-1}\dif_\Sigma g\big) = - \dif t\wedge\dif x\,\Res_{w=0}\frac{\dif w}{w^4}\tr\big(f(g^{-1}\p_x g + wg^{-1}\p_tg)\big) \\
	&= \dif t\wedge\dif x\,\bigg(\frac{2}{3}vr\p_xr - \frac{2}{3}r^2\p_xv + v\p_tr - r\p_tv\bigg)\,.
	\eea
	We conclude that
	\be \label{eq:KdV-derivative} - \int_\Sigma\Res_{w=0}\frac{\dif w}{w^2}\wedge\tr\big(a_0\wedge g^{-1}\dif_\Sigma g\big) = \int_\Sigma\dif t\wedge\dif x\,\big(2r\p_tv + r^2\p_xv\big)\ee
	This leaves
	\be
	\int_\Sigma\Res_{w=0}\frac{\dif w}{w^2}\wedge\tr\big(a_0\wedge g^{-1}\Delta\cL g\big)
	= - \int_\Sigma\dif t\wedge\dif x\Res_{w=0}\frac{\dif w}{w^3}\wedge\tr\big((f-ew)g^{-1}A_tg\big)\,.
	\ee
	Conjugation by $g$ gives
	\be g^{-1}A_tg = \exp(-w\ad_B)A_t = A_t - w\ad_BA_t + \frac{w^2}{2}\ad_B^2A_t + \mrm{O}(w^3)\,, \ee
	and using $B = ve - rh$, $A_t = - \p_xve + \p_xrh + cf$ we find
	\bea \ad_BA_t = 2(r\p_xv - v\p_xr)e + cvh + 2crf\,,\qquad \tr(f\ad_B^2A_t) = - 2cv^2 - 4r(r\p_xv - v\p_xr)\,. \eea
	Hence the final term contributes
	\be \label{eq:KdV-DeltaL} - \int_\Sigma\Res_{w=0}\frac{\dif w}{w^2}\wedge\tr\big(a_0\wedge g^{-1}\Delta\cL g\big) = \int_\Sigma\dif t\wedge\dif x\,\big(c(2r-v^2) - 3r^2\p_xv\big)\,. \ee
	
	\paragraph{Collecting terms.} Putting together equations \eqref{eq:KdV-bulk-evaluated}, \eqref{eq:KdV-derivative} and \eqref{eq:KdV-DeltaL}
	\be S_\text{4dCS}[a] = \int_\Sigma\dif t\wedge\dif x\,\big(4r\p_tv + 2(\p_xr)^2 - 4r^2\p_xv + c(2r - \p_xv - v^2)\big)\,. \ee
	At this stage we recall that $r = \tfrac{1}{2}(\p_xv + v^2)$ is simply a notational shorthand, whereas $c$ is an independent field which is left unfixed when compactifying to $\Sigma$.  Substituting $r$ into the above we can see that $c$ completely drops out of the 2d action, and we're left with
	\be S_\text{4dCS}[a] = \int_\Sigma\dif t\wedge\dif x\,\bigg(2\p_xv\p_tv + \frac{1}{2}(\p_x^2v)^2 + 2v\p_xv\p_x^2v - (\p_xv)^3 + 2v^2\p_tv - v^4\p_xv\bigg)\,. \ee
	The final two terms are exact, since $v^2\p_tv = \tfrac{1}{3}\p_t(v^3)$ and $v^4\p_xv = \frac{1}{5}\p_x(v^5)$.  The third and fourth combine, since
	\be 2v\p_xv\p_x^2v = v\p_x\big((\p_xv)^2\big) = \p_x\big(v(\p_xv)^2\big) - (\p_xv)^3\,, \ee
	leaving
	\be S_\text{4dCS}[a] = \int_\Sigma\dif t\wedge\dif x\,\bigg(2\p_xv\p_tv + \frac{1}{2}(\p_x^2v)^2 - 2(\p_xv)^3\bigg) = S_\text{KdV}[v] \ee
	as claimed.
\end{proof}


\subsection{Real Structure} \label{subsec:KdV-reality}

In order to recover the KdV equation with $u$ (and $v$) real we impose equivariance of $a$ under simultaneously pulling back by the anti-holomorphic involution
\be \sigma:M\to M\,,\qquad (x,t,w,\bar w)\mapsto(x,t,\bar w,w) \ee
and acting by elementwise complex conjugation
\be \tau:\fsl_2(\bbC)\to\fsl_2(\bbC)\,,\qquad S \mapsto \bar S\,. \ee
Gauge transformations are similarly required to be invariant under pullback by $\sigma$ and the corresponding anti-holomorphic involution of $\gSL_2(\bbC)$.  The background $a_0$ and boundary conditions \eqref{eq:KdV-bcs} respect this real structure.  Under this assumption the action is real and the potential $v$ is a real function on $\Sigma$.


\section{Generalising the Boundary Conditions} \label{sec:gen}

In sections \ref{sec:NLS} and \ref{sec:KdV} we defined specific singular boundary conditions for 4d Chern--Simons theory with gauge group $G = \gSL_2(\bbC)$ and spectral data $(C,\omega) = (\CP^1,\dif z)$.  We found that compactifying to $\Sigma$ produced variational formulations of NLS and KdV respectively.

In this section we'll see that these belong to a family of consistent singular boundary conditions defined using the data of a conformal Galilean structure on $\Sigma$, i.e., a rank-one nilpotent map $N:T^*\Sigma\to T^*\Sigma$.  All introduce a second order pole in the partial connection at $z=\infty$.  Through a combination of space-time diffeomorphisms and gauge transformations we'll find that, under suitable non-degeneracy assumptions, any boundary condition in this family is locally equivalent to either the NLS or KdV case.

This result shows that the two cases treated in detail in sections \ref{sec:NLS} and \ref{sec:KdV} are to a limited extent exhaustive.  Of course, there remains the freedom to change the order of singularity at $z=\infty$, the gauge group and the Riemann surface with meromorphic one-form.  One could also attempt to define singular boundary conditions using different geometric data.


\subsection{A Family of Singular Boundary Conditions}

We begin by generalising the boundary conditions introduced in sections \ref{subsec:NLS-bcs} and \ref{subsec:KdV-bcs} to a general smooth surface $\Sigma$ equipped with a rank-one map $N:T^*\Sigma\to T^*\Sigma$ obeying
\be N^2 = 0\,. \ee
The image of $N$ determines a line subbundle of the co-tangent bundle $L\subset T^*\Sigma$, and a corresponding line subbundle of the tangent bundle annihilating it $L^\perp\subset T\Sigma$.  Locally there exist normal co-ordinates $(t,x)$ on $\Sigma$ such that $N = \dif t\otimes\p_x$, for which $L = \la\dif t\ra_\bbR$ and $L^\perp = \la\p_x\ra_\bbR$.

We fix a singular partial connection $\alpha$ on $M$ whose space-time components are required to have second order poles as $w\to0$ and are subject to
\be \label{eq:gen-alpha} (w\mathrm{id}_\Sigma + N)\ip\alpha = \mrm{O}(1)\quad\text{as}\quad w\to0\,. \ee
Crucially we demand a non-vanishing second-order pole everywhere.  Here $\ip$ denotes the operation of contraction and wedge product with vector field valued forms.  We also require that $\alpha$ solves the equations of motion away from $w=0$, so that on $M_\eps$
\be \label{eq:gen-alpha-flat} \dift\alpha + \frac{1}{2}[\alpha,\alpha] = 0\,. \ee
The constraint \eqref{eq:gen-alpha} together with the requirement that the second order pole be non-vanishing implies that the complexified image of $T\Sigma$ under
\be \cQ = \Res_{w=0}w\alpha \in \Gamma(L\otimes\Ad_P) \ee
is a rank-one sub-bundle $\ft\subset\Ad_P|_{\Sigma_0}$.  Let $\fs = \ft^\perp$ be the orthogonal complement of $\ft$ with respect to the trace pairing $\tr$; it obeys $[\ft,\fs]\subseteq\fs$.  Notice that $\ft$ may be isotropic, in which case $\ft\subset\fs$ and $\ft,\fs$ are not complementary subspaces.  We shall impose a condition on the $\bar w$ component of $\alpha$
\be \label{eq:gen-wbar-alpha} \alpha_{\bar w}\in w\ft + \mrm{O}(w^2)\quad\text{as}\quad w\to0\,. \ee
Real structures on $\alpha$ take the same form as in sections \ref{subsec:NLS-reality} and \ref{subsec:KdV-reality}, that is,
\be \label{eq:gen-reality} \sigma^*\alpha = \tau(\alpha) \ee
for $\sigma:M\to M$ fixing $\Sigma$ and acting as $w\to\bar w$ on $\CP^1$, and $\tau$ an anti-holomorphic involution of $\fsl_2(\bbC)$ fixing either $\fsu(2),\fsu(1,1)$ or $\fsl_2(\bbR)$.

In local normal co-ordinates $(t,x)$, for which $N = \dif t\otimes\p_x$, we have $N\ip\dif x = \dif t$.  The condition \eqref{eq:gen-alpha} then reads
\be w\alpha_t + \alpha_x = \mrm{O}(1)\,,\qquad w\alpha_x = \mrm{O}(1)\quad\text{as}\quad w\to0\,. \ee
This means that $\alpha_x$ has at worst a simple pole at $w=0$, and the residue cancels that of $\alpha_t$ at its second order pole.

Our boundary conditions on fluctuations $\beta$ around the background $\alpha$ are expressible solely in terms of the line bundle $L$ and sub-bundles $\ft,\fs$.  We require that $\beta$ is non-singular as $w\to0$ and its space-time components obey
\be \label{eq:gen-bcs} \beta_\Sigma\in L\otimes\fs + \mrm{O}(w)\,,\qquad(w\mathrm{id}_\Sigma + N)\ip \beta \in wL\otimes\ft + \mrm{O}(w^2)\quad\text{as}\quad w\to0\,. \ee
These boundary conditions are isotropic for the 4d Chern--Simons pre-symplectic form \eqref{eq:symplectic-form}.  On its $\bar w$ components, and on infinitesimal gauge transformations, we require that
\be \label{eq:gen-wbar-bcs} \eps\in w\ft + \mrm{O}(w^2)\,,\qquad\beta_{\bar w}\in w\ft + \mrm{O}(w^2)\quad\text{as}\quad w\to0\,. \ee
Infinitesimal gauge transformations close, and they preserve the boundary conditions on fluctuations.  With these boundary conditions the 4d Chern--Simons functional can be defined as in equations \eqref{eq:NLS-4dCS+source} and \eqref{eq:KdV-4dCS+source} by
\be \label{eq:gen-reg-action} S_\text{4dCS}[\beta;\alpha] = \frac{\im}{2\pi}\int_M\frac{\dif w}{w^2}\wedge\tr\bigg(\beta\wedge\Dift_\alpha\beta + \frac{1}{3}\beta\wedge[\beta,\beta]\bigg) \ee
where $\Dift_\alpha = \dift + \ad_\alpha$ and the right hand side is to be understood using the $\eps$ regulator.

In the case $(\Sigma,N) = (\bbR^2,\dif t\otimes\p_x)$ and
\be \alpha = a_0^\mrm{NLS} = \frac{1}{w^2}\im h(\dif t- w\dif x) \ee
the boundary conditions \eqref{eq:gen-bcs} and \eqref{eq:gen-wbar-bcs} on $\beta$ reproduce the boundary conditions \eqref{eq:NLS-bcs}, \eqref{eq:NLS-eps-bcs} and \eqref{eq:NLS-wbar-bcs} on $b$ leading to NLS.  Similarly for $(\Sigma,N) = (\bbR^2,\dif t\otimes\p_x)$ and
\be \alpha = a_0^\mrm{KdV} = - \frac{1}{w^2}(f - we)(\dif t- w\dif x) \ee
the boundary conditions \eqref{eq:gen-bcs} and \eqref{eq:gen-wbar-bcs} on $\beta$ specialise to the boundary conditions \eqref{eq:KdV-bcs}, \eqref{eq:KdV-eps-bcs} and \eqref{eq:KdV-wbar-bcs} on $b$ which give KdV.


\subsection{Standard Form}

We will now see that, under suitable assumptions, the boundary conditions \eqref{eq:gen-bcs} and \eqref{eq:gen-wbar-bcs} are equivalent to those for NLS or KdV.  To show this we adopt the following strategy:  First we exploit a result of Mason and Sparling \cite{Mason:1989kk} which uses a combination of space-time diffeomorphisms and non-singular but otherwise unrestricted gauge transformations to bring $\alpha$ into the from of the NLS or KdV Lax for an on-shell solution.  Second we perform a further boundary condition violating gauge transformation to shift the data of this on-shell solution from the background into the fluctuations.  This leaves the Lax of the zero solution to KdV or NLS, $a_0$, with fluctuations obeying the required boundary conditions.

We will assume that $\alpha$ is holomorphically trivial on the $\CP^1$ fibres of $M$, so that there exists a gauge transformation $\theta_1:M\to\gSL_2(\bbC)$ such that
\be \cL_0 = \theta_1(\dift + \alpha)\theta_1^{-1} \ee
with $\cL_{0,\bar w} = 0$.  The flatness condition on $\alpha$ implies that $\cL_0$ is meromorphic in $w$ with at worst double poles as $w\to0$.  In local normal co-ordinates, the boundary conditions \eqref{eq:gen-alpha} fix
\be \label{eq:gen-L0} \cL_0 = \bigg(\frac{1}{w^2}Q - \frac{1}{w}P + A_t\bigg)\dif t + \bigg( - \frac{1}{w}Q + A_x\bigg)\dif x\,. \ee
for $A\in\Omega^1(\bbR^2;\fsl_2(\bbC))$ a connection one-form and $Q,P\in C^\infty(\bbR^2;\fsl_2(\bbC))$ adjoint valued scalars.  Since $\alpha$ obeys the equation of motion \eqref{eq:gen-alpha-flat}, $\cL_0$ obeys the zero-curvature equations.  An important consequence is that
\be \p_x\tr(Q^2) = 0\,,\qquad \p_x\tr(PQ) = \frac{1}{2}\p_t\tr(Q^2)\,. \ee

Now let us quote the necessary theorem of Mason and Sparling, specialised to the case of attractive NLS.  A completely analogous theorem applies in the repulsive case, and the subsequent argument follows identically.

\begin{theorem}[Mason, Sparling \cite{Mason:1989kk}] \label{thm:Mason-beta}
	Suppose $\cL_0$ is a Lax connection with the form given in equation \eqref{eq:gen-L0} and obeying the zero-curvature equation.  Further assume the equivariance condition $\sigma^*\cL_0 = \tau(\cL_0)$ for $\tau$ the anti-holomorphic involution fixing $\fsu(2)$.  Then if $\tr(Q^2)$ is nowhere vanishing, there exist local normal co-ordinates and a spectral parameter independent gauge transformation $\theta_2:\bbR^2\to\gSU(2)$ such that
	\be \cL_0^\prime = \theta_2(\dif_\Sigma + \cL_0)\theta_2^{-1} \ee
	where
	\bea \label{eq:gen-NLS-Lax}
	&\cL_{0,t}^\prime + \frac{1}{w}\cL_{0,x}^\prime = \frac{\im}{2}\begin{pmatrix} - |\psi_0|^2 & \p_x\psi_0 \\ \p_x\bar\psi_0 & |\psi_0|^2 \end{pmatrix}\,,\qquad \cL_{0,x}^\prime = - \frac{1}{w}\begin{pmatrix} \im & 0 \\ 0 & -\im \end{pmatrix} + \begin{pmatrix} 0 & \psi_0 \\ -\bar\psi_0 & 0 \end{pmatrix}
	\eea
	with $\psi_0$ solving the attractive NLS equation.
\end{theorem}

We will exploit this theorem to prove the following result:

\begin{claim} \label{claim:beta}
	Fix a background singular partial connection $\alpha$ with a double pole in its space-time components as $w\to0$ which obeys the boundary conditions \eqref{eq:gen-alpha}, $\gSU(2)$ reality conditions \eqref{eq:gen-reality} and equation of motion \eqref{eq:gen-alpha-flat}.  Further assume that it's holomorphically trivial when restricted to the $\CP^1$ fibres of $M$, and that $\tr(\cQ\otimes\cQ)$ is nowhere vanishing.
	
	Then, locally in $\Sigma$, the variational problem defined by the functional \eqref{eq:gen-reg-action} subject to the boundary conditions \eqref{eq:gen-bcs} and \eqref{eq:gen-wbar-bcs} is equivalent to that determined by the functional \eqref{eq:NLS-4dCS+source} for the background $a_0 = \im h(\dif t - w\dif x)/w^2$ and subject to the boundary conditions \eqref{eq:NLS-bcs}, \eqref{eq:NLS-eps-bcs} and \eqref{eq:NLS-wbar-bcs}.
\end{claim}

\begin{proof}
Under the above assumptions on $\alpha$ we can use Theorem \ref{thm:Mason-beta} to construct a spectral parameter dependent gauge transformation $\theta_2\theta_1:M\to\gSL_2(\bbC)$ taking $\alpha$ to $\cL_0^\prime$ with the form \eqref{eq:gen-NLS-Lax} in suitable normal co-ordinates.

\paragraph{Engineering the gauge transformation.}  Next we need to eliminate the $\psi_0$ dependence from $\cL_0^\prime$, to bring it into the standard form $a_0 = \im h(\dif t - w\dif x)/w^2$ plus a fluctuation obeying the NLS boundary conditions.  Fortunately we've already seen how to do this in section \ref{subsec:NLS-Lax}.  Let
\be X_0 = - \frac{\im}{2}(\psi_0 e + \bar\psi_0 f)\,, \ee
and define $\theta_3 = \exp( - wB(|w|)X_0)$ using the bump function $B:[0,\infty)\to[0,1]$ taking the value $1$ in a neighbourhood of $w=0$ and $0$ for sufficiently large arguments.  Then our proof of Claim \ref{claim:NLS-Lax} shows that
\be \cL_0^\prime = \theta_3^{-1}(\dift + a_0 + b_0)\theta_3 \ee
for $b_0$ obeying the NLS boundary conditions \eqref{eq:NLS-bcs} and \eqref{eq:NLS-wbar-bcs}.  The complete gauge transformation $\theta = \theta_3\theta_2\theta_1$ therefore brings a general background $\alpha$ into the standard form $a_0$ plus a fluctuation obeying the NLS boundary conditions.

\paragraph{Transforming the boundary conditions.}  We will now perform the change of variables
\be \beta\mapsto b = \theta(\dift + \alpha + \beta)\theta^{-1} - a_0 = b_0 + \theta\beta\theta^{-1} \ee
at the level of off-shell fields.  Notice that conjugation by $\theta_0 = \theta|_{w=0}$ rotates $\ft$ to $\fh = \la h\ra_\bbC$ and $\fs$ to $\fm = \la e,f\ra_\bbC$.  Furthermore, the condition \eqref{eq:gen-wbar-alpha} ensures that $\log\theta = \log\theta_0 + \mrm{O}(w)$, so that the general boundary conditions \eqref{eq:gen-bcs} and \eqref{eq:gen-wbar-bcs} on $\beta$ rotate into the NLS boundary conditions \eqref{eq:NLS-bcs} and \eqref{eq:NLS-wbar-bcs} on $\theta\beta\theta^{-1}$.  Since $b_0$ also obeys the NLS conditions, so does $b$.

\paragraph{Comparing functionals.}  Finally we need to show that the regulated action \eqref{eq:gen-reg-action} coincides with \eqref{eq:NLS-4dCS+source}, at least up to a background-dependent constant.  4d Chern--Simons theory is topological in the $\Sigma$ direction away from $\Sigma_0$, ensuring that the transformation to convenient normal co-ordinates preserves the action.  Furthermore, the identity
\be S_\mrm{4dCS}[\beta;\alpha] = S_\mrm{4dCS}[\theta\beta\theta^{-1};a_0 + b_0] \ee
holds at the level of the integrand, because $\Dift_{a_0+b_0}(\theta\beta\theta^{-1}) = \theta(\Dift_\alpha\beta)\theta^{-1}$.    All that remains is evaluating the difference
\bea
&S_\mrm{4dCS}[b_0 + \theta \beta\theta^{-1};a_0] - S_\mrm{4dCS}[\theta\beta\theta^{-1};a_0 + b_0] \\
&= \frac{\im}{2\pi}\int_M\frac{\dif w}{w^2}\wedge\big(\CS(b_0 + \theta\beta\theta^{-1};a_0) - \CS(\theta\beta\theta^{-1},a_0 + b_0)\big)\,,
\eea
where
\be \CS(b;a_0) = \tr\bigg(b\wedge\Dift_0b + \frac{1}{3}b\wedge[b,b]\bigg)\,. \ee
We have
\be \CS(b_0 + \theta\beta\theta^{-1};a_0) - \CS(\theta\beta\theta^{-1},a_0 + b_0) = \CS(b_0;a_0) - \dift\tr\big(b_0\wedge\theta\beta\theta^{-1})\big)\,. \ee
The first term depends only on the background $\alpha$, and the second is exact.  It does not contribute a boundary term, because both $b_0$ and $\theta\beta\theta^{-1}$ obey the NLS boundary conditions which are isotropic.  In full we have
\be S_\mrm{4dCS}[\beta;\alpha] = S_\mrm{4dCS}[b;a_0] - S_\mrm{4dCS}[b_0,a_0]\,. \ee
The variational problems therefore agree, and the functionals differ by a background dependent constant.  In fact Claim \ref{claim:NLS-compactify} shows that the constant $S_\mrm{4dCS}[b_0,a_0]$ is the on-shell action of the background NLS solution $\psi_0$.

\end{proof}

In the case of KdV the relevant counterpart of Theorem \ref{thm:Mason-beta} is:

\begin{theorem}[Mason, Sparling \cite{Mason:1989kk}] \label{thm:Mason-alpha}
	Suppose $\cL_0$ is a Lax connection with the form given in equation \eqref{eq:gen-L0} and obeying the zero-curvature equation.  Further assume the equivariance condition $\sigma^*\cL_0 = \tau(\cL_0)$ for $\tau$ the anti-holomorphic involution fixing $\fsl_2(\bbR)$.  If $\tr(Q^2) = 0$ and $\tr(PQ)$ is positive, then there exist local normal co-ordinates and a spectral parameter independent gauge transformation $\theta_2:\bbR^2\to\gSL_2(\bbR)$ such that
	\be \cL_0^\prime = \theta_2(\dif_\Sigma + \cL_0)\theta_2^{-1} \ee
	where
	\bea
	&\cL_{0,t}^\prime + \frac{1}{w}\cL_{0,x}^\prime = \begin{pmatrix} \frac{1}{2}\p_x^2v_0 + v_0\p_xv_0 & -\p_xv_0 \\ c_0 & - \frac{1}{2}\p_x^2v_0 - v_0\p_xv_0 \end{pmatrix}\,, \\
	&\cL_{0,x}^\prime = \frac{1}{w}\begin{pmatrix} 0 & 0 \\ 1 & 0 \end{pmatrix} + \begin{pmatrix} v_0 & -1 \\ \p_xv_0 + v_0^2 & - v_0 \end{pmatrix}
	\eea
	for $u_0 = \p_xv_0$ solving the KdV equation and
	\be c_0 = - \p_tv_0 + \frac{1}{2}\p_x^3v_0 + 2(\p_xv_0)^2 + v_0\p_x^2v_0 + v_0^2\p_xv_0\,. \ee
\end{theorem}

Defining $\cP = \Res_{w=0}\alpha$, we have the following analogue of claim \ref{claim:beta}.

\begin{claim} \label{claim:alpha}
	Fix a singular partial connection $\alpha$ with non-vanishing double pole in its space-time components as $w\to0$ which obeys the boundary conditions \eqref{eq:gen-alpha}, $\gSL_2(\bbR)$ reality conditions \eqref{eq:gen-reality} and equation of motion \eqref{eq:gen-alpha-flat}.  Further assume that it's holomorphically trivial when restricted to the $\CP^1$ fibres of $M$, and that $\tr(\cQ\otimes\cQ) = 0$ but $\tr(\cP\otimes\cQ)$ is positive.
	
	Then, locally in $\Sigma$, the variational problem defined by the functional \eqref{eq:gen-reg-action} subject to the boundary conditions \eqref{eq:gen-bcs} and \eqref{eq:gen-wbar-bcs} is equivalent to that determined by the functional \eqref{eq:KdV-4dCS+source} for the background $a_0 = -(f-we)(\dif t - w\dif x)/w^2$ and subject to the boundary conditions \eqref{eq:KdV-bcs}, \eqref{eq:KdV-eps-bcs} and \eqref{eq:KdV-wbar-bcs}.
\end{claim}

The proof takes a similar form as for Claim \ref{claim:beta}.


\subsection*{Acknowledgements}

It is a pleasure to thank Kevin Costello, Lionel Mason, David Skinner, Beno\^{\i}t Vicedo and Lewis Cole for useful conversations, and in particular Kevin Costello, Lionel Mason and Beno\^{\i}t Vicedo for comments on a draft version of this note.  My work is funded by the Simons Collaboration on Celestial Holography.  GPT-6 Astra was used at several stages of this project.  Of course, the author takes full responsibility for all claims and arguments in the paper.

Results appearing in this work were obtained during the `Simons Programme on Twistor Theory and its Applications' hosted by the Mathematical Institute of the Polish Academy of Sciences.  This programme was supported by the Simons Foundation grant SFI-MPS-T-Institutes-00010825, and by the State Treasury funds as part of a task commissioned by the Minister of Science and Higher Education under the project Organization of the Simons Semesters at the Banach Center - New Energies in 2026-2028 (MNiSW/2025/DAP/491).


\begin{appendix}
	
\section{Gauge Fixing} \label{app:gauge-fixing}

In this Appendix we show that:

\begin{claim}
Under the assumption of holomorphic triviality, gauge transformations $k:M\to\gSL_2(\bbC)$ obeying $\log k\in\cI_\fh$ in a neighbourhood of $w=0$ can be used to bring $a_{\bar w}\in\cI_h$ into the form
\be \label{app:trivialize} a_{\bar w} = g^{-1}\p_{\bar w}g \ee
where $g = \exp(wX)$ in a neighbourhood of $w=0$.  Here $X\in C^\infty(\Sigma;\fm)$ with $\fm = \la e,f\ra_\bbC\subset\fsl_2(\bbC)$.
\end{claim}

\begin{proof}
For simplicity we'll use the fundamental matrix representation of $\fsl_2(\bbC)$.  Holomorphic triviality implies that we can write \eqref{app:trivialize} for $g:M\to\gSL_2(\bbC)$ with $g|_{w=0} = \mathrm{id}$. Rearranging \eqref{app:trivialize} gives
\be \label{app:frame} \p_{\bar w}g = ga_{\bar w} \in wC^\infty(M;\op{Mat}_{2\times 2}(\bbC))\,, \ee
so that $g = \mathrm{id} + wG$ for $G\in C^\infty(M;\op{Mat}_{2\times 2}(\bbC))$.  We write $a_{\bar w} = wC + w^2D$ for $C\in C^\infty(M;\fh)$ and $D\in C^\infty(M;\fm)$, and substitute back into equation \eqref{app:frame} to get
\be \label{app:intermediate} \p_{\bar w}G = (\mathrm{id} + wG)(C + wD) = C + wE \ee
where $E = D + GC + wGD\in C^\infty(M;\op{Mat}_{2\times 2}(\bbC))$.  Equation \eqref{app:intermediate} is solved by
\be \label{app:crucial} G = X + H + wF \ee
where $H\in C^\infty(M;\fh)$, $F\in C^\infty(M;\op{Mat}_{2\times 2}(\bbC))$ and $X\in C^\infty(\Sigma;\fm)$.  The crucial point here is that the $\fm$ components of $G$ differ from $X$ by an expression divisible by $w$.  In a neighbourhood of $w=0$ we may choose
\be \label{app:gauge-transformation} k = e^{-wX}g\,, \ee
which we extend to the whole of $M$ using a bump function $B:[0,\infty)\to[0,1]$ taking the value $1$ in a neighbourhood of $0$ and $0$ for sufficiently large arguments as $k = \exp(-wB(|w|)X)g$.  From equation \eqref{app:crucial} the gauge transformation $k$ obeys
\be \log k = wH + \mrm{O}(w^2)\quad\text{as}\quad w\to0 \ee
for $H\in C^\infty(M;\fh)$, or equivalently $\log k\in\cI_h$ in a neighbourhood of $w=0$.  Finally, under the gauge transformation \eqref{app:gauge-transformation} $g\mapsto\exp(wX)$ near $w=0$ with $X\in C^\infty(\Sigma;\fm)$ as claimed.
\end{proof}

An identical argument shows that gauge transformations $k:M\to\gSL_2(\bbC)$ obeying $\log k\in\cI_{\fn_-}$ in a neighbourhood of $w=0$ can be used to bring $a_{\bar w}\in\cI_{\fn_-}$ into the form
\be  a_{\bar w} = g^{-1}\p_{\bar w}g \ee
where $g = \exp(wB)$ in a neighbourhood of $w=0$.  Here $B\in C^\infty(\Sigma;\fb^+)$ with $\fb^+ = \la e,h\ra_\bbC\subset\fsl_2(\bbC)$.

\end{appendix}

\bibliographystyle{JHEP}
\bibliography{references}


\end{document}